\documentclass[amsmath,amssymb,amsbsy,pra,twocolumn,superscriptaddress
]{revtex4-1}

\usepackage{graphicx}
\usepackage{braket}
\usepackage{bm}
\usepackage{color}
\usepackage{ulem}
\usepackage{mathtools}
\usepackage{amsfonts}
\usepackage[breaklinks,colorlinks=true,linkcolor=blue,urlcolor=blue,citecolor=magenta]{hyperref}

\def\ze{\mathbb{Z}}

\begin{document}
\title{Majorana edge magnetization in the Kitaev honeycomb model}

\author{Tomonari Mizoguchi}
\email{mizoguchi@rhodia.ph.tsukuba.ac.jp}
\affiliation{Department of Physics, University of Tsukuba, Tsukuba, Ibaraki 305-8571, Japan}
\author{Tohru Koma}
\email{tohru.koma@gakushuin.ac.jp}
\affiliation{Department of Physics, Gakushuin University, Mejiro, Toshima-ku, Tokyo 171-8588, Japan}

\date{\today}
\begin{abstract}
We propose an approach to detect the peculiarity of Majorana fermions at the edges of Kitaev magnets. 
As is well known, a pair of Majorana edge modes is realized when a single complex fermion splits into real and imaginary parts which 
are, respectively, localized at the left and right edges of a sample magnet. 
Reflecting both of this peculiarity of the Majorana fermions 
and the ground-state degeneracy 
caused by the existence of the Majorana edge zero modes, 
the spins at the edges of the sample magnet are expected to behave 
as a peculiar \lq\lq free" spin which exhibits a unidirectional magnetization 
without any transverse magnetization 
when applied a sufficiently weak external magnetic field. 
For the Kitaev honeycomb model, we obtain the expression of 
the Majorana edge magnetization 
by relying on standard techniques to diagonalize a free fermion Hamiltonian. 
The magnetization profile thus obtained indeed shows the expected behavior. 
We also elucidate the relation between the Majorana edge flat band and the bulk 
winding number from a weak topological point of view. 
\end{abstract}

\maketitle

\section{Introduction}
Kitaev introduced a seminal quantum spin model~\cite{Kitaev2006}
which realizes the desired properties of quantum spin 
liquids~\cite{Anderson1973,Balents2010}, whose study was initiated by Anderson. 
More precisely, Kitaev's model is defined on the honeycomb 
lattice, and mapped to a free Majorana ferminon model. 
In consequence, the model is exactly solvable, and 
shows short-range spin correlations~\cite{Baskaran2007}. 
Although Kitaev's model is fairly artificial, some materials, 
e.g., $A_2$IrO$_3$ ($A=$ Na, Li)~\cite{Singh2010,Singh2012} and $\alpha$-RuCl$_3$~\cite{Plumb2014}, are expected to exhibit 
very similar properties to those of the Kitaev honeycomb model~\cite{Jackeli2009,Rau2014,Trebst2017,Winter2017,Hermanns2018}. 
In particular, detecting the evidence of the Majorana 
fermions is one of the central issues in condensed matter 
physics~\cite{Sandilands2015,Banerjee2016,Glamazda2016,Banerjee2017,Do2017,Kasahara2018,Kasahara2018_2}. 

On the other hand, the Kitaev honeycomb model with open boundaries 
shows many Majorana zero modes at the edges~\cite{Thakurathi2014}.
In fact, the zero modes form a flat band. This is nothing but a consequence of the weak topological 
character~\cite{Halperin1987,Montambaux1990,Kohmoto1992,Fu2007,Moore2007,Roy2009} of the model. 
In fact, the celebrated bulk-edge correspondence~\cite{Halperin1982,Hatsugai1993}
ensures the relation between the number of the Majorana edge zero modes 
and the winding number for the bulk Hamiltonian when the Fermi energy, 
which equals zero in the present system, 
lies in the spectral or mobility gap of the Hamiltonian. 
Interestingly, similar Majorana edge flat bands were 
found to appear also in the gapless regime of the Hamiltonian, 
depending on the geometry of the edges~\cite{Thakurathi2014}.  

As is well known, a pair of Majorana edge modes appears   
when a single complex fermion splits into real and imaginary parts which 
are, respectively, localized at the left and right edges of the system. 
Therefore, a single Majorana fermion has only a real degree of freedom 
as an internal degree of freedom. 
Reflecting this peculiarity of the Majorana fermions 
and the ground-state degeneracy 
caused by the existence of the Majorana edge zero modes, 
the spins at the edges of the Kitaev honeycomb model are expected to behave 
as a peculiar \lq\lq free" spin which exhibits a unidirectional magnetization 
without any transverse magnetization 
when applied a sufficiently weak external magnetic field. 
Thus, the edge magnetization is expected to serve as a probe to detect 
the signature of the Majorana edge modes. 

\begin{figure}[b]
\begin{center}
\includegraphics[width= 0.9\linewidth]{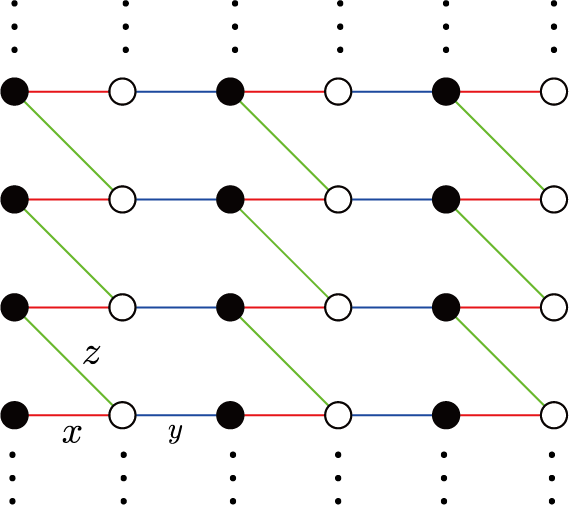}
\caption{The Kitaev honeycomb model with edges in the horizontal direction
and the periodic boundary condition in the vertical direction.
Red, blue, and green bonds represent 
$x$- $y$- and $z$-bonds, respectively. 
Black (white) circles denote 
the sites with an odd (even) $\ell$. }
\label{Fig1}
\end{center}
\end{figure}
In the present paper, we derive the expression of the edge magnetization 
in the bulk gapped regime of the Kitaev honeycomb model on the cylinder geometry 
with the open boundary condition, by using standard techniques 
to diagonalize a free-fermion Hamiltonian. 
As a result, we show that the profile of the edge magnetization indeed 
exhibits the desired properties. 
More precisely, the magnetization is perfectly 
unidirectional, namely, only a single component of the expectation value of 
the three-component spin operator can be non-vanishing, while the rest two components 
must be vanishing. This is nothing but a consequence of the fractionalization of 
a single spin into two independent Majorana fermions. 

We also elucidate the relation between the Majorana edge flat band and 
the weak topological character of the Kitaev honeycomb model. The latter is 
characterized by the winding number for the Hamiltonian on the infinitely-long cylinder with a finite radius. Due to the topological 
character, the Majorana edge flat band can be proved to be stable against 
disorders which preserve the symmetry of the Hamiltonian. Further, we can 
expect that the Majorana edge flat band is stable against 
additional interactions. 

The subject treated in the present paper is similar to that in Refs.~\onlinecite{Willans2011} and \onlinecite{Sreejith2016}, i.e., 
the local magnetization at the vacancy sites.
In both of the two cases with vacancies and edges, the localized zero-energy modes play an important role in the emergence of local magnetization.
However, we stress that the emergence of the edge magnetization is a consequence of 
the bulk-edge correspondence in the Kitaev magnet as a weak topological material. 

The rest of this paper is organized as follows. 
In Sec. \ref{sec:model}, 
we give the precise definition of the Kitaev's honeycomb model 
with the cylinder geometry, which is periodic in the vertical direction and 
has two edges in the horizontal direction. The model is mapped to a fermion model  
by using the Jordan-Wigner transformation. 
In Sec. \ref{sec:edge}, 
for the model with the cylinder geometry, 
we derive zero-energy edge states in the momentum space. 
Further, from these states, we construct Wannier orbitals, which are useful for calculating the local quantities.
In Sec. \ref{sec:magnetization}, we compute the edge magnetization by explicitly calculating the expectation value 
of the spin operator by using the Wannier orbitals.  
Section \ref{sec:discussion} is devoted to discussions which include the stability of the Majorana edge magnetization, comparisons with previous works and possible experimental realizations.
The summary of the present paper is given in Sec. \ref{sec:summary}.
Appendix \ref{sec:app} is devoted to 
a proof of the bulk-edge correspondence for all of the classes which show a topologically nontrivial 
index in one dimension.  The correspondence clarifies the relation between the Majorana edge flat band 
and the bulk winding number in the present system.
In Appendix \ref{Sec:appB}, we show an alternative approach to calculate the edge magnetization.
In Appendix \ref{sec:appC}, we show how to construct the edge zero modes in 
the bond-disordered Kitaev honeycomb model. 

\section{Kitaev honeycomb model and Jordan-Wigner transformation
\label{sec:model}}
We consider the spin-$1/2$ Kitaev magnet on the honeycomb lattice. 
The Hamiltonian is given by 
\begin{equation}
H = \sum_{\gamma\in\{x,y,z\}}\; \sum_{\langle i,j\rangle\in {\cal B}_\gamma} J_\gamma \sigma_i^\gamma
\sigma_j^\gamma, 
\label{eq:hamspin}
\end{equation}
where $\sigma_i^\gamma$ is the $\gamma$ component of the Pauli 
matrices for $\gamma=x,y,z$ at the site $i$ in the honeycomb lattice 
shown in Fig.~\ref{Fig1}, and ${\cal B}_\gamma$ is the set of all the bonds $\langle i,j\rangle$ 
(pairs of nearest neighbor two sites $i,j$ in the honeycomb lattice) 
with the type $\gamma$ whose three types, $x,y,z$, are, respectively, 
denoted by three colors, red, blue and green, in Fig.~\ref{Fig1}; 
$J_\gamma$ is the corresponding exchange integral which is a real parameter. 
We impose the open boundary condition in the horizontal direction and 
the periodic boundary condition in the vertical direction in Fig.~\ref{Fig1}. 
Namely, we consider the cylinder geometry with two zigzag edges.
Clearly, one can notice that the sites, denoted by black and white circles in Fig.~\ref{Fig1}, 
are placed on the two-dimensional square lattice. 
Therefore, we can label a site $i$ by $i=(\ell,m)$ 
with two positive integers, $\ell$ and $m$, which satisfy 
$1\le \ell\le 2L_x$ and $1\le m\le L_y$ with the length $2L_x$ of the cylinder 
and the length $L_y$ of the circumference of the cylinder.   

The Hamiltonian of Eq.~(\ref{eq:hamspin})  
can be transformed to the Majorana Hamiltonian (\ref{eq:Hamiltonian_Majorana}) below 
by using the Jordan-Wigner transformation as follows. 
Following Refs. \onlinecite{Feng2007,Yao2007,Chen2007,Chen2008}, 
we first introduce a fermion operator $a_{(\ell,m)}$ at the site $(\ell,m)$ 
such that the Pauli matrices are represented as 
\begin{equation}
\sigma_{(\ell,m)}^+ = 2 a_{(\ell,m)}e^{i\pi\hat{\theta}_{(\ell,m)}}, \label{eq:JW_sigmax}
\end{equation}
\begin{equation}
\sigma_{(\ell,m)}^- = 2e^{i\pi\hat{\theta}_{(\ell,m)}} a^\dagger_{(\ell,m)}, \label{eq:JW_sigmay}
\end{equation}
\begin{equation}
\sigma_{(\ell,m)}^z = (-1)^\ell\left[ 2a_{(\ell,m)}^{\dagger}a_{(\ell,m)} -1 \right], \label{eq:JW_sigmaz}
\end{equation}
where $\sigma_i^{\pm} = \sigma_i^x \pm i\sigma_i^y$, and 
\begin{eqnarray}
\hat{\theta}_{(\ell,m)} =\sum_{m^\prime < m }\sum_{\ell^\prime = 1}^{2L_x} 
a^{\dagger}_{(\ell^\prime,m^\prime)}a_{(\ell^\prime,m^\prime)} 
+ \sum_{\ell^\prime < \ell }a^{\dagger}_{(\ell^\prime,m)}a_{(\ell^\prime,m)}.\nonumber \\
\end{eqnarray}
Further, by introducing the Majorana fermions, $c_i$ and $d_i$, as~\cite{Feng2007,Yao2007,Chen2007,Chen2008}
\begin{eqnarray}
c_{(\ell,m)} = &  i \left[a_{(\ell,m)}^\dagger -a_{(\ell,m)}\right], \nonumber\\
d_{(\ell,m)} = & a_{(\ell,m)}^\dagger + a_{(\ell,m)}\ \ \mbox{if $\ell$ is odd}, \label{eq:Majoranaodd}
\end{eqnarray}
and 
\begin{eqnarray}
c_{(\ell,m)} = &a_{(\ell,m)}^\dagger + a_{(\ell,m)}, \nonumber\\
d_{(\ell,m)} =  & i\left[a_{(\ell,m)}^\dagger - a_{(\ell,m)}\right]\ \ \mbox{if $\ell$ is even}, \label{eq:Majoranaeven}
\end{eqnarray}
the Hamiltonian of Eq. (\ref{eq:hamspin}) can be written as 
\begin{eqnarray}
H &=&  i J_x \sum_{\ell = 1}^{L_x} \sum_{m = 1}^{L_y} c_{(2\ell -1,m)} c_{(2\ell,m)} \nonumber\\
 &+&  i J_y\sum_{\ell = 1}^{L_x-1} \sum_{m = 1}^{L_y} c_{(2\ell,m)} c_{(2\ell  +1,m)} \nonumber\\
&+& J_z \sum_{\ell = 1}^{L_x} \sum_{m = 1}^{L_y} c_{(2\ell,m)} c_{(2\ell-1,m+1)} 
d_{(2\ell,m)} d_{(2\ell -1,m  +1)}.  \nonumber\\
\label{eq:Hamiltonian_Majorana}
\end{eqnarray}

Since any pairs of $z$-bonds do not share a site, $d_{(2\ell,m)} d_{(2\ell -1,m  +1)}$ commute with the Hamiltonian.
This allows us to replace this operator with the c-number $d_{(2\ell,m)} d_{(2\ell -1,m  +1)} = \pm i$,
which acts as a phase factor that leads to an effective flux for itinerant $c$-Majorana fermions.
For the system on the torus, Lieb's theorem ensures that the ground state is obtained when the flux configuration is uniform~\cite{Lieb1994}.
Clearly, the uniform flux can be realized by setting all the eigenvalues of $d_{(2\ell,m)} d_{(2\ell -1,m  +1)}$
to be the same for a given $m$~\cite{Feng2007}.
For the cylinder geometry which we consider here, we have numerically confirmed 
that the ground state is obtained in the same configuration of the eigenvalue of $d_{(2\ell,m)} d_{(2\ell -1,m  +1)}$.
In the following, we set $d_{(2\ell,m)} d_{(2\ell -1,m  +1)} = + i$ for every $(\ell,m)$,
and we represent the corresponding wavefunction as $\ket{\Phi^{(0)}_{\mathrm{flux}}}$. 

In this case, the Hamiltonian is written in terms of the $c$-Majorana fermion as
\begin{align}
 H =  & i J_x \sum_{\ell = 1}^{L_x} \sum_{m = 1}^{L_y} c_{(2\ell -1,m)} c_{(2\ell,m)} \notag \\
 + &  i J_y\sum_{\ell = 1}^{L_x-1} \sum_{m = 1}^{L_y} c_{(2\ell ,m)} c_{(2\ell + 1,m)} \notag \\
 + & i J_z \sum_{\ell = 1}^{L_x} \sum_{m = 1}^{L_y} c_{(2\ell,m)} c_{(2\ell-1,m+1)}. \label{eq:hami}
 \end{align}
 
In order to elucidate the topological nature of the Hamiltonian of 
Eq. (\ref{eq:hami}) in the next section, we need to know in advance what symmetries the Hamiltonian has.
To do this, we first rewrite it in a Bogoliubov-de Gennes (BdG) form.
To be specific, we introduce complex fermions $\alpha_{(\ell,m)}$ that is a combination of 
$c_{(2\ell -1 , m)}$ and $c_{(2\ell,m)}$: 
\begin{equation}
\alpha_{(\ell,m) } =  \frac{1}{2} [c_{(2\ell -1 , m)} + i c_{(2\ell,m)} ],  \label{eq:def_alpha}
\end{equation}
with $\ell = 1,2, \cdots, L_x$, $m = 1,2, \cdots, L_y$. 
Using the complex fermions $\alpha_{(\ell,m)}$, we can rewrite the Hamiltonian as 
\begin{widetext}
\begin{align}
 H = & J_x \sum_{\ell = 1}^{L_x} \sum_{m = 1}^{L_y}  \left[ \alpha^\dagger_{(\ell,m)} \alpha_{(\ell,m)} -\alpha_{(\ell,m)} \alpha^\dagger_{(\ell,m)} \right] \notag \\
 + &  J_y \sum_{\ell = 1}^{L_x-1} \sum_{m = 1}^{L_y} \left[\alpha_{(\ell,m)} \alpha_{(\ell +1,m)} - \alpha^\dagger_{(\ell,m)} \alpha_{(\ell + 1,m)}
 + \alpha_{(\ell,m)} \alpha^\dagger_{(\ell+1,m)} -\alpha^\dagger_{(\ell,m)} \alpha^\dagger_{(\ell+1,m)} \right] \notag \\
 + & J_z \sum_{\ell = 1}^{L_x} \sum_{m = 1}^{L_y} \left[ \alpha_{(\ell,m)}\alpha_{(\ell,m+1)} - \alpha^\dagger_{(\ell,m)} \alpha_{(\ell,m + 1)} + \alpha_{(\ell,m)}\alpha^\dagger_{(\ell,m+1)} -\alpha^\dagger_{(\ell,m)} \alpha^\dagger_{(\ell,m+1)} 
 \right]. \label{eq:BDG}
\end{align}
\end{widetext}
Further, we write $\tilde{\alpha}_i^-=\alpha_i$ and 
$\tilde{\alpha}_i^+=\alpha_i^\dagger$ with $i=(\ell,m)$.
For the superscript $\zeta=\pm$ of the new fermion operators $\tilde{\alpha}_i^\pm$, we define $\overline{\zeta}$ as 
$\overline{\zeta}=-$ if $\zeta=+$ and $\overline{\zeta}=+$ if $\zeta=-$. 
Then, the Hamiltonian can be written in the form, 
\begin{equation}
H=\sum_{\xi,\eta = \pm }\sum_{i,j}\tilde{\alpha}_i^{\overline{\xi}}{\cal H}_{ij}^{\xi\eta}\tilde{\alpha}_j^\eta,
\end{equation}
where ${\cal H}$ is the complex-valued matrix, i.e., the single-body Hamiltonian. 
Let $\varphi$ be a single-body wavefunction whose components are given by $\varphi_i^\zeta$ 
with $i=(\ell,m)$. 
The explicit action of the Hamiltonian ${\cal H}$ for the wavefunction $\varphi$ is written as 
\begin{eqnarray}
({\cal H}\varphi)_{(\ell,m)} =&J_x\tau^z\varphi_{(\ell,m)}  \nonumber\\
- &\frac{J_y}{2}(\tau^z-i\tau^y) \varphi_{(\ell-1,m)}-\frac{J_y}{2}(\tau^z+i\tau^y)\varphi_{(\ell+1,m)} \nonumber\\
- & \frac{J_z}{2}(\tau^z-i\tau^y) \varphi_{(\ell,m-1)}-\frac{J_z}{2}(\tau^z+i\tau^y)\varphi_{(\ell,m+1)}, \nonumber\\ \label{eq:eigenvalueeq}
\end{eqnarray}
where $\tau^\gamma$ is the $\gamma$ component of the Pauli matrices which act on 
the internal degree of freedom denoted by the superscript $\zeta=\pm$. 

The form in Eq. (\ref{eq:eigenvalueeq}) enables us to find simple forms of the symmetry operations.  
Firstly, it is invariant under the time reversal transformation given by the simple complex conjugate, since the Hamiltonian ${\cal H}$ is real symmetric. 
Secondly, under the transformation given by $\tau^x$, the Hamiltonian ${\cal H}$ changes its sign.  
Thus, it has the chiral symmetry, too. 
Finally, as is well known, the combination of time reversal and chiral symmetries gives the particle-hole symmetry. 
In consequence, the classification class of the Hamiltonian as a topological material 
is BDI class~\cite{AZ,Schnyder2008,Ryu2010}. 
The topological invariant is an integer-valued winding number 
for the system on the infinity-long strip. 
Further, if this number is nonvanishing, then there appears an edge zero mode 
in the system with an edge by virtue of the bulk-edge correspondence,
as we will see in the next section. 

It should also be remarked that, for the armchair edge, 
the corresponding $c$-Majorana Hamiltonian has neither the chiral nor the time-reversal symmetries,
thus the corresponding topological class becomes D class.
The loss of symmetries originates from the fact that 
the complex fermion of Eq. (\ref{eq:def_alpha}) is not compatible with the armchair edge,
and we need to use a different set of complex fermions in order to obtain the BdG Hamiltonian. 
Therefore, due to the $\mathbb{Z}_2$ character of D class, the Majorana edge flat band 
cannot be expected to be topologically stable against generic perturbations. 
In other words, only the parity of the number of the zero-energy edge modes is topologically protected 
in D class. (See Appendix~\ref{sec:app}.)

 \section{Majorana edge flat band in A$_{\mathrm{y}}$ phase \label{sec:edge}}
 \subsection{Solution of BdG equation in momentum space}
\begin{figure}[h]
\begin{center}
\includegraphics[width=\linewidth]{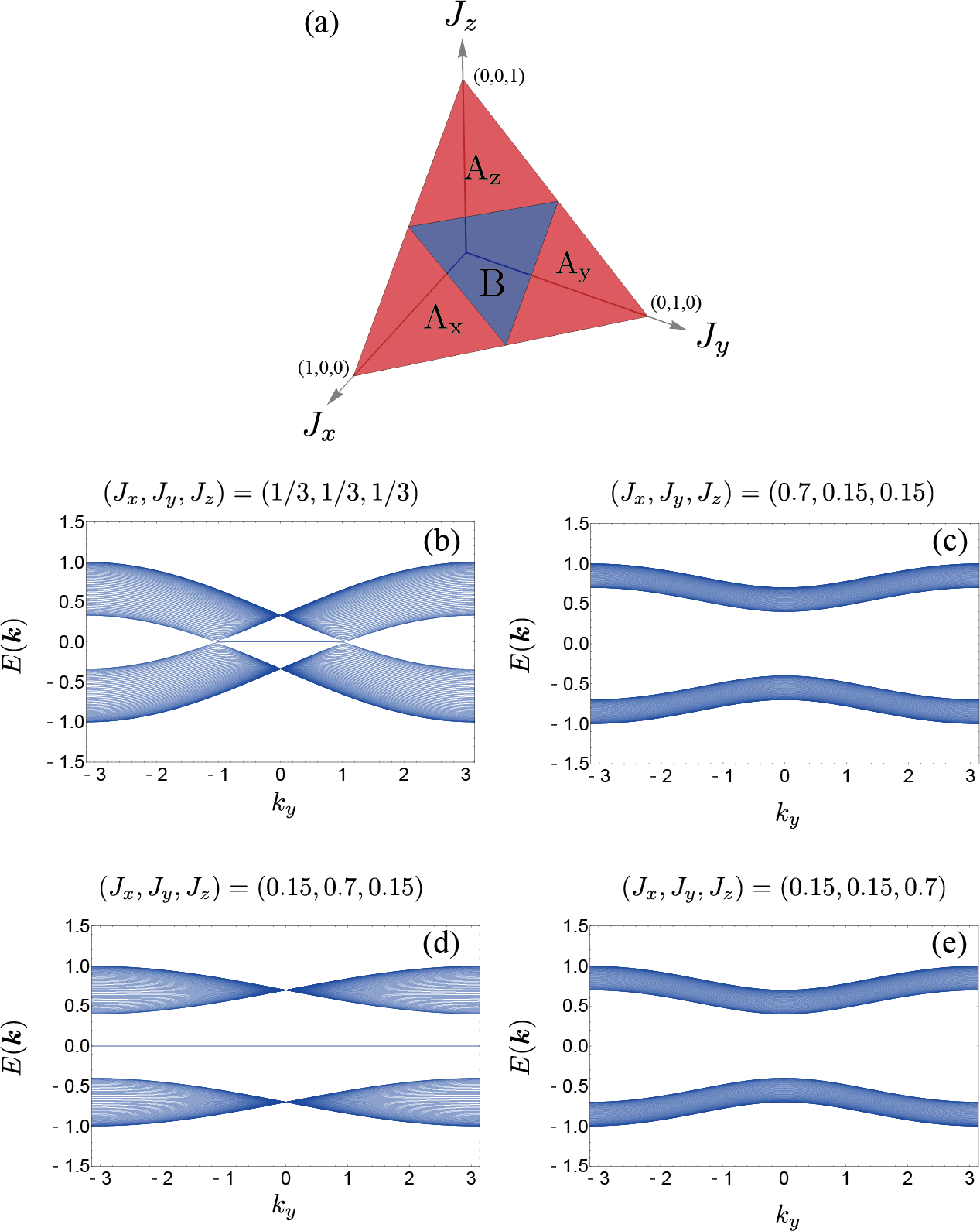}
\caption{(a) The phase diagram of the Kitaev honeycomb model with $J_x + J_y +J_z = 1$.
Dispersion relations of itinerant Majorana fermions 
for (b) B phase, (c) A$_{\rm x}$ phase, 
(d) A$_{\rm y}$ phase, and (e) A$_{\rm z}$ phase. 
}
\label{Fig2}
\end{center}
\end{figure}
Using the Fourier transform in the vertical direction,
\begin{equation}
\alpha_{\ell, k_y} =\frac{1}{\sqrt{L_y}}  \sum_{m = 1}^{L_y}  e^{i k_y m } \alpha_{(\ell,m)}, \label{eq:alpha_fourier}
\end{equation}
we obtain the BdG form of the Hamiltonian in the momentum space:
\begin{equation}
H = \sum_{k_y} \Psi^{\dagger}(k_y) \hat{h}(k_y) \Psi(k_y),   \label{eq:majorana_hamiltonian}
\end{equation}
where
\begin{equation}
\Psi(k_y)  = (\alpha_{1,k_y},  \cdots ,
\alpha_{L_x, k_y} ,\alpha^{\dagger}_{1,- k_y},  \cdots, \alpha^{\dagger}_{L_x, -k_y} )^{\mathrm{T}},
\end{equation}
and
\begin{equation}
\hat{h}(k_y) = 
\left(
\begin{array}{cc}
\hat{h}_0 (k_y) & \hat{\Delta} (k_y) \\
\hat{\Delta} ^\dagger(k_y) & - \hat{h}_0 (-k_y) \\
\end{array}
\right),
\end{equation} 
with
\begin{equation}
[\hat{h}_0 (k_y)]_{\ell, \ell^\prime} = J_x-J_z \cos k_y \delta_{\ell, \ell^\prime}  -\frac{J_y}{2} (\delta_{\ell,\ell^\prime-1} +\delta_{\ell,\ell^\prime+1}),  \label{h0}
\end{equation} 
\begin{equation}
[\hat{\Delta} (k_y)]_{\ell, \ell^\prime} = - i J_z \sin k_y \delta_{\ell, \ell^\prime}  -\frac{J_y}{2} (\delta_{\ell, \ell^\prime-1} - \delta_{\ell, \ell^\prime+1}). \label{delta}
\end{equation} 

The energy spectra for four different phases~\cite{Kitaev2006} 
are shown in Fig.~\ref{Fig2}.
Among four phases, B phase has gapless spectrum in the bulk 
and the flat band with zero energy at the edges. 
Only A$_\mathrm{y}$ phase shows both of the gapped spectrum in the bulk 
and the flat band with zero energy at the edges for the present geometry of the edges.
This is associated with the topological nature of the Hamiltonian of Eq. (\ref{eq:majorana_hamiltonian}).

The origin of the zero energy modes can be understood by considering the special case. 
Namely, let us assume that $J_z=0$. 
Then, the model becomes the independent $L_y$ chains. For a large $J_y$, 
each chain shows the winding number $1$. Therefore, the $L_y$ chains 
give the total winding number $L_y$ due to the additivity of the index. 
As is well known, the homotopy argument guarantees that 
when varying the model parameters continuously, i.e., restoring the interchain 
coupling $J_z$, the topological invariant, 
or the winding number, does not change as long as the spectral or mobility gap of the Hamiltonian 
does not close. Thus, the winding number always takes the value $L_y$ in 
the whole regime A$_{\rm y}$. 
Due to the bulk-edge correspondence, this implies that there appear 
zero-energy edge modes whose number is at least $L_y$ (per edge). 
This is nothing but a flat edge band. 

In the following, we consider A$_\mathrm{y}$ phase, 
and derive the wave function of zero-energy modes by directly solving the BdG equation. 
Since the zero modes are doubly-degenerate at each $k_y$, 
we label them as $\gamma^{(\nu)}_{k_y}$, $\nu = 1,2$. 
Each zero mode can be expanded by $\alpha$ of (\ref{eq:alpha_fourier}) as 
\begin{equation}
\gamma^{(\nu)}_{k_y} = \sum_{\ell = 1}^{L_x} u^{(\nu)}_{\ell,k_y} \alpha_{\ell,k_y} + v^{(\nu)}_{\ell,k_y} \alpha^\dagger_{\ell,- k_y}, \label{eq:def_gamma}
\end{equation}
with the coefficients, $u^{(\nu)}_{\ell,k_y}$ and $v^{(\nu)}_{\ell,k_y}$.
The energy eigenvalue equation of 
the coefficients, $u^{(\nu)}_{\ell,k_y}$ and $v^{(\nu)}_{\ell,k_y}$,
at zero energy is given by
\begin{equation}
\hat{h}(k_y) 
\left(
\begin{array}{c}
u^{(\nu)}_{1,k_y} \\
\vdots\\
u^{(\nu)}_{L_x,k_y} \\
v^{(\nu)}_{1,k_y} \\
\vdots\\
v^{(\nu)}_{L_x,k_y} \\
\end{array}
\right) = 0. \label{eq:eigen1}
\end{equation}
From Eqs. (\ref{h0}) and (\ref{delta}), Eq. (\ref{eq:eigen1}) can be rewritten as 
\begin{eqnarray}
K_1 u^{(\nu)}_{\ell,k_y} - & K_2  v^{(\nu)}_{\ell,k_y} - \frac{J_y}{2} (u^{(\nu)}_{\ell-1,k_y} - v^{(\nu)}_{\ell-1,k_y}) \nonumber \\
- &\frac{J_y}{2} (u^{(\nu)}_{\ell + 1,k_y} + v^{(\nu)}_{\ell + 1,k_y}) = 0, \label{eq:eigen2}
\end{eqnarray} 
\begin{eqnarray}
-K_1 v^{(\nu)}_{\ell,k_y}  + & K_2 u^{(\nu)}_{\ell,k_y} - \frac{J_y}{2}  (u^{(\nu)}_{\ell-1,k_y} - v^{(\nu)}_{\ell-1,k_y}) \nonumber \\
 + &\frac{J_y}{2} (u^{(\nu)}_{\ell + 1,k_y} + v^{(\nu)}_{\ell + 1,k_y}) = 0,\label{eq:eigen3}
\end{eqnarray} 
where $K_1 = J_x -J_z \cos k_y$, and $K_2 = i J_z \sin k_y$.
To solve this, we introduce $\xi^{(\nu)}_{\ell,k_y} \equiv u^{(\nu)}_{\ell,k_y} - v^{(\nu)}_{\ell,k_y}$ and $\zeta^{(\nu)}_{\ell,k_y} \equiv u^{(\nu)}_{\ell,k_y} + v^{(\nu)}_{\ell,k_y}$. 
By adding and subtracting the two equations, (\ref{eq:eigen2}) and (\ref{eq:eigen3}), we obtain 
\begin{equation}
(K_1 + K_2) \xi^{(\nu)}_{\ell,k_y}   - J_y  \xi^{(\nu)}_{\ell-1,k_y} =0, \label{eq:eigen4}
\end{equation}
and
\begin{equation}
(K_1 - K_2) \zeta^{(\nu)}_{\ell,k_y}   - J_y  \zeta^{(\nu)}_{\ell + 1,k_y} =0.  \label{eq:eigen5}
\end{equation}

Since we consider A$_\mathrm{y}$ phase, i.e., a large $J_y$, 
we treat the case that the parameters of the Hamiltonian satisfy 
the condition,
\begin{equation}
\left| \frac{K_1 \pm K_2}{J_y} \right| < 1,
\end{equation}
or, equivalently,
\begin{equation}
\sqrt{J_x^2 + J_z^2 -2J_xJ_z \cos k_y} < |J_y|.  \label{eq:condition}
\end{equation}
As we will see below, this condition is enough to find the edge zero modes~\cite{Thakurathi2014}. 
Further, we consider the system with 
a sufficiently large length $L_x$ so that the exponential correction 
in the length $L_x$ can be ignored.  
Then, Eq. (\ref{eq:eigen4}) has no left-edge 
solution because of the open boundary condition $\xi_{0,k_y}=0$, but it has 
the right-edge solution,  
\begin{eqnarray}
\xi^{(1)}_{\ell,k_y}  = \left( \frac{K_1 + K_2}{J_y} \right)^{L_x- \ell } \xi^{(1)}_{L_x,k_y}.\nonumber \\
\end{eqnarray}
Similarly, Eq. (\ref{eq:eigen5})  has no right-edge solution by the boundary condition $\zeta_{L_x+1,k_y}=0$, 
but it shows the left-edge solution, 
\begin{eqnarray}
\zeta^{(2)}_{\ell,k_y}  = \left( \frac{K_1 - K_2}{J_y} \right)^{\ell-1} \zeta^{(2)}_{1,k_y}.
\end{eqnarray}
Since all of $\xi_{\ell,k_y}^{(2)}$ are vanishing for the latter case, 
i.e., the left-edge mode, we obtain
\begin{eqnarray}
u^{(2)}_{\ell,k_y} =
v^{(2)}_{\ell,k_y}  = 
\frac{\left( \frac{J_x -J_ze^{ik_y}}{J_y} \right)^{\ell-1}}{N(k_y)}, \label{eq:u2}
\end{eqnarray}
where $N(k_y)$ is the normalization factor as a function of $k_y$. 

\subsection{Wannier orbitals}
The flatness of the Majorana edge band enables us to construct Wannier orbitals which are localized in the real-space. 
Namely, we define $\gamma^{(\nu)}_n$ as 
\begin{widetext}
\begin{align}
\gamma^{(\nu)}_n = & \frac{1}{\sqrt{L_y} } \sum_{k_y} e^{-i k_y n} \gamma^{(\nu)}_{k_y}
=  \frac{1}{\sqrt{L_y} }\sum_{k_y} e^{-i k_y n}\left( \sum_{\ell= 1}^{L_x} u_{\ell,k_y}^{(\nu)} \alpha_{\ell, k_y} +  v_{\ell,k_y}^{(\nu)} \alpha^\dagger_{\ell,-k_y}  \right)  \notag \\
=  &  \frac{1}{L_y} \sum_{k_y} e^{-i k_y n}\left\{ \sum_{\ell= 1}^{L_x} \left[ \sum_{m= 1}^{L_y} e^{i k_y m} \left(  u^{(\nu)}_{\ell,k_y}  \alpha_{(\ell,m)}
+ v^{(\nu)}_{\ell,k_y}  \alpha^\dagger_{(\ell,m)} \right) 
 \right]  \right\}. \label{eq:wannier}
\end{align}
\end{widetext}
Introducing the coefficients $U^{(\nu)}_{n} (\ell,m)$ and $V^{(\nu)}_{n}(\ell,m)$ as  
\begin{equation}
U^{(\nu)}_{n} (\ell,m) \equiv \frac{1}{L_y} \sum_{k_y}  e^{i k_y (m -n) } u^{(\nu)}_{\ell,k_y}, \label{eq:U}
\end{equation}                     
\begin{equation}
V^{(\nu)}_{n}(\ell,m) \equiv \frac{1}{L_y} \sum_{k_y}  e^{i k_y (m- n ) } v^{(\nu)}_{\ell,k_y},\label{eq:V}
\end{equation}
$\gamma^{(\nu)}_n$ is written as 
\begin{align}
\gamma^{(\nu)}_n = \sum_{\ell = 1}^{L_x} \sum_{m = 1}^{L_y}  
U^{(\nu)}_{n} (\ell,m) \alpha_{(\ell,m)} +V^{(\nu)}_{n}(\ell,m)  \alpha^\dagger_{(\ell,m)}    .
\end{align}

\begin{figure}[t]
\begin{center}
\includegraphics[width= 0.95\linewidth]{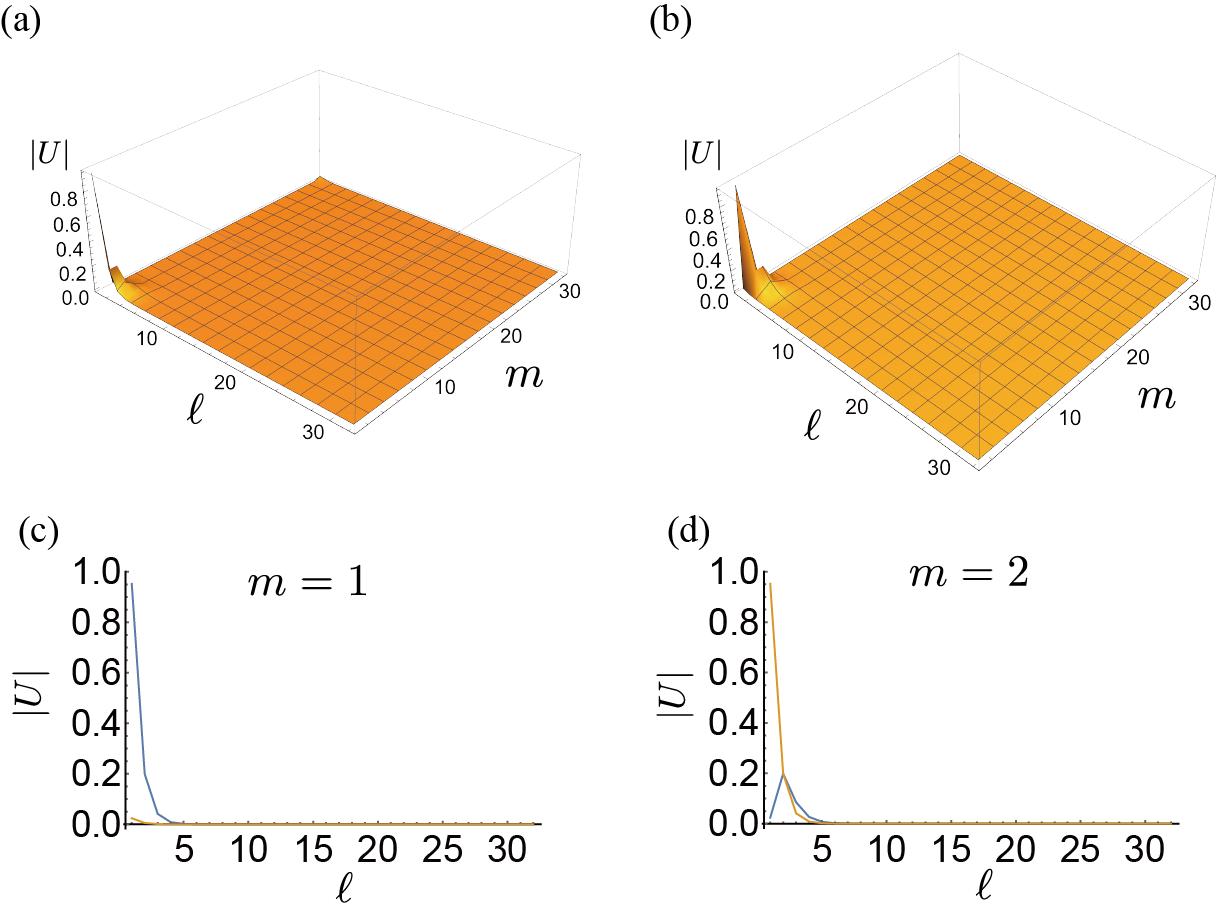}
\caption{The absolute value of the coefficient $U^{(2)}_{n} (\ell,m)$ for (a) $n=1$ and (b) $n=2$
for the system with $L_x = L_y = 32$. 
(c) and (d) are for fixed $m$ to $m=1$ and $m=2$, respectively. 
Blue and orange lines denote $n=1$ and $n=2$, respectively.
The exchange parameters used here are $(J_x,J_y,J_z) = (0.15,0.7,0.15)$.
}
\label{Fig3}
\end{center}
\end{figure}

In the following, we will consider only the left edge mode, i.e., the case with $\nu=2$,  
because we can deal with the case of the right edge mode in the same way. 
Clearly, from the condition $u_{\ell,k_y}^{(\nu)} = v_{\ell,k_y}^{(\nu)}$ in the case with $\nu=2$ and the definitions, (\ref{eq:U}) and (\ref{eq:V}), of $U^{(2)}_{n} (\ell,m)$ and $V^{(2)}_{n} (\ell,m)$, 
one has $U^{(2)}_{n} (\ell,m) = V^{(2)}_{n} (\ell,m)$. 
Therefore, the Wannier function of (\ref{eq:wannier}) with $\nu=2$ can be written as
\begin{eqnarray}
\gamma^{(2)}_n = &  \sum_{\ell = 1}^{L_x} 
\sum_{m}^{L_y}  U^{(2)}_{n} (\ell,m) [  \alpha_{(\ell,m)} + \alpha^\dagger_{(\ell,m)}] \nonumber \\
 = & \sum_{\ell = 1}^{L_x} \sum_{m = 1}^{L_y}   U^{(2)}_{n} (\ell,m) c_{(2\ell-1,m)}, \label{eq:gamma_majorana}
\end{eqnarray}
where we have used the relation (\ref{eq:def_alpha}).
From Eq. (\ref{eq:U}) and the explicit form (\ref{eq:u2}) of $u_{\ell,k_y}^{(2)}$, 
one notices $U^{(2)}_{n} (\ell,m)$ is real.
Therefore, $\gamma^{(2)}_n$ of (\ref{eq:gamma_majorana}) satisfies the Majorana condition, $[\gamma^{(2)}_n]^\dagger= \gamma^{(2)}_n$, 
i.e., it is a Majorana fermion. 
We choose the normalization factor $N(k_y)$ so that $ \gamma^{(2)}_n$ 
satisfies
$\{ \gamma^{(2)}_n, \gamma^{(2)}_{n^\prime} \} = 2 \delta_{n,n^\prime}$,
which leads to
\begin{equation}
\sum_{\ell = 1}^{L_x} \sum_{m = 1}^{L_y} [U^{(2)}_{n} (\ell,m)]^2 = 1. 
\end{equation} 
In Fig.~\ref{Fig3}, we show $|U^{(2)}_{n} (\ell,m)|$ for $n=1$ and $2$. 
We can see that 
the distribution of 
$|U^{(2)}_{n} (\ell,m)|$ 
is concentrated at 
$(\ell,m) = (1,n)$, 
and rapidly decays with the distance.

\section{Edge magnetization \label{sec:magnetization}}
Now we compute the edge magnetization at the left edge.
More precisely, we calculate the expectation value $\frac{1}{L_y}\sum_{m=1}^{L_y}\langle \sigma_{(1 ,m) }^{\gamma} \rangle $, $\gamma = x,y,z$, 
where $\langle  O\rangle$ stands for the expectation value of the operator $O$ with respect to 
a ground state. 

\subsection{Warm-up: single-site magnetization}
Before calculating the edge magnetization, it is instructive to calculate the 
magnetization at $(\ell,m) = (1,1)$.
The spin operators $\sigma_{(1,1)}^\gamma$ are written in terms of the Majorana fermions, $c_i, d_i$, 
by using Eqs. (\ref{eq:JW_sigmax})-(\ref{eq:JW_sigmaz}) and Eq. (\ref{eq:Majoranaodd}), as 
\begin{equation}
\sigma^x_{(1,1)} = d_{(1,1)},  \label{eq:sigmax_majorana}
\end{equation}
\begin{equation}
\sigma^y_{(1,1)} = c_{(1,1)},  \label{eq:sigmay_majorana}
\end{equation}
\begin{equation}
\sigma^z_{(1,1)} = id_{(1,1)}c_{(1,1)}.  \label{eq:sigmaz_majorana}
\end{equation}
Clearly, both of the two operators, $\sigma^x_{(1,1)}$ and $\sigma^z_{(1,1)}$, contain 
the single Majorana fermion $d_{(1,1)}$. We recall the well known fact~\cite{Feng2007,Yao2007,Chen2007,Chen2008} that 
the gauge-field sector of the ground state is the eigenstate of pairs of the $d$-Majorana fermions, 
$d_{(2 \ell,m)}d_{(2 \ell -1,m+1)}$, whose eigenvalue acts as an effective phase for the itinerant $c$-Majorana 
fermions. This fact also means that the fermion parity of the $d$-Majorana fermions is a good quantum number. 
Combining these observations with the fact that the ground state has a uniform flux, 
one notices that a Majorana excitation created by an odd number of the operators $d_i$ above 
the ground state is gapped in the present A$_{\rm y}$ phase.
This implies that the expectation values of $\sigma^x_{(1,1)}$ and $\sigma^z_{(1,1)}$ with respect to 
the ground state must be vanishing. 
On the other hand, the Majorana operator $c_{(1,1)}$ in the right-hand side of Eq. (\ref{eq:sigmay_majorana}) 
creates a Majorana edge zero mode above a ground state as 
we will see in Eq.~(\ref{eq:ex1}) below. 
Therefore, we can expect that only the expectation value of $\sigma^y_{(1,1)}$ is nonvanishing.
We note that the vanishing of the expectation value for $x$- and $z$- components occurs for arbitrary $(1,m)$ with $m=2,3, \cdots$,
thus the entire edge magnetization also shows the unidirectional nature.

To calculate the expectation value of $\sigma_{(1 ,1)}^{y}$ with respect to a ground-state vector, 
we recall that the ground state of the $c$-Majorana fermion sector is degenerate
due to the zero modes which form the flat band at the edge. 
Let us consider a form of the ground state
which is given by
\begin{equation}
|\mathrm{GS}_{\mathrm{s}} \rangle = (s + t \gamma_1^{ (2)})
|0, \Phi^{(0)}_{\mathrm{flux}} \rangle, 
\label{eq:GS}
\end{equation}
where the complex numbers, $s$ and $t$, satisfy $|s|^2 + |t|^2 = 1$ for the normalization 
$\langle{\mathrm{GS}_{\mathrm{s}}}|{\mathrm{GS}_{\mathrm{s}}}\rangle=1$, 
and $| 0, \Phi^{(0)}_{\mathrm{flux}}  \rangle$ is the total ground state 
including the flux sector for $d$-Majorana fermions,
whose negative energy levels are all occupied by the usual 
complex fermions which diagonalize the Hamiltonian.
Then, the expectation value of $\sigma_{(1,1)}^{y}$ with respect to $|\rm{GS}_s\rangle$ is written as 
\begin{eqnarray}
\langle \mathrm{GS}_{\mathrm{s}}|\sigma_{(1,1)}^{y} |\mathrm{GS}_{\mathrm{s}}\rangle 
= \langle{0}| 
(s^\ast + t^\ast \gamma_1^{(2)}) c_{(1,1)}(s+ t\gamma_1^{ (2)}) |{0}\rangle.\nonumber\\
\label{eq:ex1}
\end{eqnarray}
In order to calculate this right-hand side, we expand the operator $c_{(1,m)}$ in terms of the fermions 
which diagonalize the Hamiltonian as 
\begin{equation}
c_{(1,m)}= W^m_1\gamma_1^{(2)} + W^m_2\gamma_2^{(2)}+\ldots +W^m_{L_y}\gamma_{L_y}^{(2)}+\ldots,   \label{eq:c_expand}
\end{equation}
where we have written $\gamma_n^{(2)}$ terms only because the rest of the terms do not contribute to 
the magnetization as we will show below. 
The coefficients $W^m_n$ are determined by the anticommutator with $\gamma_n^{(2)}$, i.e., 
\begin{equation}
\{\gamma_n^{(2)},c_{(1,m)}\}= 2W^m_n.
\end{equation}
On the other hand, from Eq. (\ref{eq:gamma_majorana}), we have 
\begin{equation}
\{c_{(1,m)},\gamma_n^{(2)}\}=2U_{n}^{(2)}(1,m). \label{eq:anticommute}
\end{equation}
Combining these equations, we have 
\begin{eqnarray}
c_{(1,m)} &=& U_{1}^{(2)}(1,m)\gamma_1^{(2)} + U_{2}^{(2)}(1,m)\gamma_2^{(2)}  \nonumber \\
&+& \ldots +U_{L_x}^{(2)}(1,m)\gamma_{L_x}^{(2)}+\ldots.     \label{eq:c_expand_2}
\end{eqnarray}
Substituting 
(\ref{eq:c_expand_2}) with $m=1$ into (\ref{eq:ex1}), we obtain 
\begin{eqnarray}
 \langle \mathrm{GS}_{\mathrm{s}}|\sigma_{(1,1)}^{y} |\mathrm{GS}_{\mathrm{s}}\rangle 
 =  (s^\ast t+st^\ast)
U_{1}^{(2)}(1,1), \label{eq:magnetization_final}
\end{eqnarray} 
where we have used the anticommutativity and the normalization $(\gamma_1^{(2)})^2=1$ for $\gamma_1^{(2)}$,
and the fermion parity conservation for $\ket{0}$.

The complex numbers, $s$ and $t$, in (\ref{eq:magnetization_final}) are determined as follows:
If we apply an infinitesimally weak magnetic field 
in the $y$-direction at the site $(1,1)$, the magnetization $\langle{\rm GS}_{\mathrm{s}}|\sigma_{(1 ,1)}^{y}|{\rm GS}_{\mathrm{s}}\rangle$ 
is maximized so as to gain the maximum Zeeman energy. 
Thus, we have to determine the coefficients $s$ and $t$ so that the magnetization is maximized.
We can easily find its maximum value of the magnetization, 
\begin{equation}
\langle{\rm GS}_{\mathrm{s}}|\sigma_{(1 ,1)}^{y}|{\rm GS}_{\mathrm{s}}\rangle = U^{(2)}_{1} (1,1). \label{eq:magnetization_app2}
\end{equation}
by choosing $s = t = \frac{1}{\sqrt{2}}$ up to the overall phase factor.
From this result, we see that,
from the expression (\ref{eq:gamma_majorana}) of $\gamma_n^{(2)}$ in terms of $c_{(2\ell-1,m)}$, 
the value of the single-site magnetization is given by the 
amplitude of the Majorana fermion $c_{(1,1)}$ in the Wannier orbital $\gamma^{(2)}_1$. 
Another key observation is that we need a linear combination of 
$\ket{0} $ and $\gamma_1^{(2)} \ket{0}$, namely the fermion parity of $c$-Majorana fermion has to be mixed, to obtain a finite magnetization, since 
the spin operator has an odd parity of $c$-Majorana fermion.

A reader might think that the uniform edge magnetization per length of the edge 
is given by the right-hand side of Eq. (\ref{eq:magnetization_app2}) because of the translational invariance of the present system 
in the vertical direction. However, this is a non-trivial problem. To see this, consider the case with 
two-site magnetization.
The natural extension of Eq. (\ref{eq:GS}) for two-site magnetization is given as 
\begin{eqnarray}
|\mathrm{GS} ^\prime \rangle = (s + t \gamma_1^{ (2)})(s + t \gamma_2^{ (2)})
|0, \Phi^{(0)}_{\mathrm{flux}} \rangle. \label{eq:gsprime}
\end{eqnarray} 
Then, by using the expression (\ref{eq:c_expand_2}) in the same way, one can compute 
the expectation value of $\sigma_{(1,1)}^y$ as follows:  
\begin{eqnarray}
&& \langle \mathrm{GS}^\prime |\sigma_{(1,1)}^{y} |\mathrm{GS}^\prime \rangle \nonumber \\
&=&  \sin 2\beta \left[ U_1^{(2)} (1,1) + \cos 2\beta U_2^{(2)} (1,1)\right],
\end{eqnarray}
where we have introduced $\beta:= \arctan \frac{t}{s}$. 
By setting $\beta = \frac{\pi}{4} + \delta \beta$ with a small $|\delta \beta|$, 
one obtains $\langle \mathrm{GS}^\prime |\sigma_{(1,1)}^{y} |\mathrm{GS}^\prime \rangle \sim U_1^{(2)}(1,1) - 2 U_2^{(2)}(1,1) \delta\beta$.
Since $U_2^{(2)}(1,1) > 0$, the expectation value of 
$\sigma_{(1,1)}^{y}$ can exceed the right-hand side of Eq. (\ref{eq:magnetization_app2})
for appropriately choosing $\delta \beta < 0 $.
Thus, the magnetization at the site (1,1) increases for the trial state (\ref{eq:gsprime}),  
while the magnetization at the neighboring site (1,2) is vanishing as follows: 
\begin{eqnarray}
&&  \langle \mathrm{GS}^\prime |\sigma_{(1,2)}^{y} |\mathrm{GS}^\prime \rangle  \nonumber \\
 &=& \langle \mathrm{GS}^\prime | c_{(1,2)} e^{i\pi \hat{\theta}_{(1,2)} } |\mathrm{GS}^\prime \rangle  \nonumber \\
 &=& 0,
\end{eqnarray}
because the operator $e^{i\pi \hat{\theta}_{(1,2)} }$ acts not only on the $c$-fermion sector but also 
the $d$-fermion sector (see the next subsection for details). 
Thus, obtaining the maximum uniform magnetization at the edge is a non-trivial problem.  
Indeed, the appropriate choice of the ground state is essential to obtain 
the edge magnetization, which we will elucidate in the next subsection.

\subsection{Edge magnetization}
Keeping these observations in mind, we move on to the calculation of the entire edge magnetization. 
To do this, we have to calculate the expectation value
$\langle \sigma_{(1 ,m) }^{y} \rangle $ with arbitrary $m$.
Again, from Eqs. (\ref{eq:JW_sigmax})-(\ref{eq:JW_sigmaz}) and Eq. (\ref{eq:Majoranaodd}), we obtain the Majorana representation of the spin operator:
\begin{equation}
\sigma^y_{(1,m)} = c_{(1,m)}e^{i \pi \hat{\theta}_{(1,m)}}. \label{eq:sigmay_majorana_2}
\end{equation}
From Eq. (\ref{eq:sigmay_majorana_2}), we see that we now have to deal with the non-local operator $e^{i \pi \hat{\theta}_{(1,m)}}$,
which is the {\it total} fermion parity operator (i.e., the fermion parity for both $c$- and $d$- Majorana fermions) below $m$-th row. 

What wavefunction do we need to choose to obtain the finite expectation value for all $m  =1,2, \cdots L_y$?
To see this, let us consider a trial wavefunction, 
\begin{equation}
|\mathrm{GS}_{\mathrm{e}} \rangle =2^{-\frac{L_y}{2}}\mathcal{S}[(1+\Gamma_1)(1+\Gamma_2)\cdots(1+\Gamma_{L_y})] |0, \Phi_{\mathrm{flux}}^{(0)}\rangle,  \label{eq:GS_edge}
\end{equation}
where 
\begin{equation}
\Gamma_n =e^{i\pi\hat{\theta}_{(1,n)}}\gamma_n^{(2)}
\end{equation}
for $n=1,2,\ldots,L_y$, and $\mathcal{S}[\cdots]$ is the spatially ordered product which is defined as follows:  
\begin{eqnarray}
\mathcal{S}[\Gamma_{r_1}\Gamma_{r_2}\cdots \Gamma_{r_N}]
= &e^{i\pi\hat{\theta}_{(1,r_1)}}e^{i\pi\hat{\theta}_{(1,r_2)}}\cdots e^{i\pi\hat{\theta}_{(1,r_N)}} \nonumber \\
\times &\gamma_{r_N}^{(2)}\cdots\gamma_{r_2}^{(2)}\gamma_{r_1}^{(2)} 
\end{eqnarray}
for $r_1,r_2,\ldots,r_N$ satisfying $r_1<r_2<\cdots <r_N$. 
Although the form of the wavefunction in Eq. (\ref{eq:GS_edge}) is seemingly complicated, one can see that 
this is a natural extension of $\ket{\rm GS_\mathrm{s}}$, namely, the factor $\frac{1}{\sqrt{2}}(1 + \gamma_1^{(2)})$ is replaced 
with a product over the entire edge, $\mathcal{S} \left[\prod_{n=1}^{L_y} \frac{1}{\sqrt{2}} (1 + \Gamma_n) \right]$.
However, there are two sharp differences between (\ref{eq:GS}) and (\ref{eq:GS_edge}).
Firstly, the expression of (\ref{eq:GS_edge})
includes not only $\gamma_n^{(2)}$ but also $e^{i\pi\hat{\theta}_{(1,n)}}$.
Since $e^{i\pi\hat{\theta}_{(1,n)}}$ acts on  both $c$-Majorana fermion sector and the flux sector, 
the flux part of $|\rm{GS}_{\mathrm{e}} \rangle$ is no longer equal to the original one. 
Nevertheless, $|\rm{GS}_{\mathrm{e}} \rangle$ is still a ground state of the Hamiltonian,
as we will show below. 
Secondly, there is a spatial ordering operator in Eq. (\ref{eq:GS_edge}).
Actually, thanks to the spacial ordering, we can show that $|\rm{GS}_e \rangle$ is indeed a ground state. 
To be more specific, one can show that $e^{i\pi\hat{\theta}_{(1,n)}}$ commute with the Hamiltonian, 
which means that for an arbitrary ground state of the Hamiltonian $\ket{\mathrm{GS}}$, $e^{i\pi\hat{\theta}_{(1,n)}}\ket{\mathrm{GS}}$ a ground state as well. 
Applying this to the terms in (\ref{eq:GS_edge}), we observe the following: since $\gamma_{r_N}^{(2)}\cdots\gamma_{r_2}^{(2)}\gamma_{r_1}^{(2)} |0, \Phi_{\mathrm{flux}}^{(0)} \rangle$ is a ground state,
so is the state $e^{i\pi\hat{\theta}_{(1,r_1)}}e^{i\pi\hat{\theta}_{(1,r_2)}}\cdots e^{i\pi\hat{\theta}_{(1,r_N)}}\gamma_{r_N}^{(2)}\cdots\gamma_{r_2}^{(2)}\gamma_{r_1}^{(2)} |0, \Phi_{\mathrm{flux}}^{(0)} \rangle$.

Let us proceed to the calculation of the expectation value 
$\langle \mathrm{GS}_{\mathrm{e} }|\sigma_{(1,m)}| \mathrm{GS}_{\mathrm{e} } \rangle = \langle \mathrm{GS}_{\mathrm{e} }|c_{(1,m)} e^{i\pi\hat{\theta}_{(1,m)}}| \mathrm{GS}_{\mathrm{e} } \rangle$.
To perform this calculation, let us recall the fact that 
the operator $e^{i\pi\hat{\theta}_{(1,m)}}$ changes the fermion parity of a $d$-Majorana-fermion 
state for a bond belonging to ${\cal B}_\gamma$. 
This leads to $\bra{\psi,\Phi_{\mathrm{flux}}^{(0)}}  e^{i\pi\hat{\theta}_{(1,m)}} \ket{\psi, \Phi_{\mathrm{flux}}^{(0)}} = 0 $
for an arbitrary choice of $\ket{\psi}$ as a wavefunciton in the $c$-fermion sector.
Thus, the non-vanishing contributions in the expectation value are given by the terms,  
\begin{widetext}
\begin{eqnarray}
2^{-L_y}\langle 0, \Phi_{\mathrm{flux}}^{(0)}|\gamma_{r_1}^{(2)}\gamma_{r_2}^{(2)}\cdots\gamma_{r_{p-1}}^{(2)}\gamma_{r_{p+1}}^{(2)}\cdots\gamma_{r_N}^{(2)}
e^{i\pi\hat{\theta}_{(1,r_N)}}\cdots e^{i\pi\hat{\theta}_{(1,r_{p-1})}}e^{i\pi\hat{\theta}_{(1,r_{p+1})}}
\cdots e^{i\pi\hat{\theta}_{(1,r_2)}}   e^{i\pi\hat{\theta}_{(1,r_1)}}\nonumber\\
\times c_{(1,m)}e^{i\pi\hat{\theta}_{(1,m)}}
e^{i\pi\hat{\theta}_{(1,r_1)}}e^{i\pi\hat{\theta}_{(1,r_2)}}\cdots e^{i\pi\hat{\theta}_{(1,r_p)}}\cdots
e^{i\pi\hat{\theta}_{(1,r_N)}}
\gamma_{r_N}^{(2)}\cdots\gamma_{r_p}^{(2)}\cdots\gamma_{r_2}^{(2)}\gamma_{r_1}^{(2)}|0, \Phi_{\mathrm{flux}}^{(0)} \rangle,
\label{expandterm} 
\end{eqnarray}
\end{widetext}
where $r_p = m$. 
We note that $\{c_{(1,m)},e^{i\pi\hat{\theta}_{(1,n)}}\}=0$ for $m<n$ and 
$[c_{(1,m)},e^{i\pi\hat{\theta}_{(1,n)}}]=0$ for $m\ge n$. These commutation relations are a consequence of the facts that 
the operator $e^{i\pi\hat{\theta}_{(1,n)}}$ is the fermion parity below the $n$-th row and that 
the fermion operator $c_{(1,m)}$ changes the fermion parity at the $m$-th row. 
Using these relations and the anticommutativity of $\gamma_n^{(2)}$, the term of (\ref{expandterm}) 
can be written as 
\begin{widetext}
\begin{eqnarray}
2^{-L_y} \langle 0|\gamma_{r_1}^{(2)}\gamma_{r_2}^{(2)}\cdots\gamma_{r_{p-1}}^{(2)}\gamma_{r_{p+1}}^{(2)}\cdots\gamma_{r_N}^{(2)}
c_{(1,m)}\gamma_m^{(2)}\gamma_{r_N}^{(2)}\cdots\gamma_{r_{p-1}}^{(2)}\gamma_{r_{p+1}}^{(2)}
\cdots\gamma_{r_2}^{(2)}\gamma_{r_1}^{(2)}|0 \rangle.
\end{eqnarray}
\end{widetext}
By adding the conjugate contribution, we have 
\begin{widetext}
\begin{eqnarray}
2^{-L_y} \langle 0|\gamma_{r_1}^{(2)}\gamma_{r_2}^{(2)}\cdots\gamma_{r_{p-1}}^{(2)}\gamma_{r_{p+1}}^{(2)}\cdots\gamma_{r_N}^{(2)}
\{c_{(1,m)},\gamma_m^{(2)}\}\gamma_{r_N}^{(2)}\cdots\gamma_{r_{p-1}}^{(2)}\gamma_{r_{p+1}}^{(2)}
\cdots\gamma_{r_2}^{(2)}\gamma_{r_1}^{(2)}|0 \rangle. \label{eq:edgemag_1}
\end{eqnarray}
\end{widetext}
From (\ref{eq:anticommute}), we find that the anticommutator in (\ref{eq:edgemag_1}) is equal to $2U^{(2)}_m(1,m)$.
Therefore, the corresponding contribution becomes 
$2U_m^{(2)}(1,m)$. 
Recalling the fact that there are $2^{L_y-1}$ choices of such $r_1, \cdots, r_N$ (notice that one of the $r_1, \cdots, r_N$ has to be equal to $m$),
we obtain $\langle \mathrm{GS}_{\mathrm{e} }|\sigma_{(1,m)}^y|\mathrm{GS}_{\mathrm{e}} \rangle=U_m^{(2)}(1,m)$. 
In the same way, one can also show that $\ket{\mathrm{GS}_{\mathrm{e}}}$ is normalized, i.e., 
$\langle \mathrm{GS}_{\mathrm{e} }|\mathrm{GS}_{\mathrm{e}}\rangle=1$. 
Consequently, we have 
\begin{eqnarray}
 \frac{1}{L_y} \sum_{m=1}^{L_y} \langle \mathrm{GS}_{\mathrm{e} } | \sigma^y_{(1,m)}|\mathrm{GS}_{\mathrm{e} }   \rangle
 = &\frac{1}{L_y} \sum_{m=1}^{L_y}U_m^{(2)}(1,m) \nonumber \\
 = & U_1^{(2)}(1,1) \label{eq:edgemag_final}
\end{eqnarray}
To obtain the final line of (\ref{eq:edgemag_final}), 
we use the relation $U_m^{(2)}(1,m) = U_1^{(2)}(1,1)$ which can be derived from (\ref{eq:U}). 
We remark that, although we cannot exclude the possibility that there is an alternative choice of the ground state 
 having a larger edge magnetization than that for $\ket{\mathrm{GS}_{\mathrm{e} }}$,
the present result of (\ref{eq:edgemag_final}) provides the lower bound of the edge magnetization under the infinitesimally small external magnetic field. 

\begin{figure}[b]
\begin{center}
\includegraphics[width=0.95\linewidth]{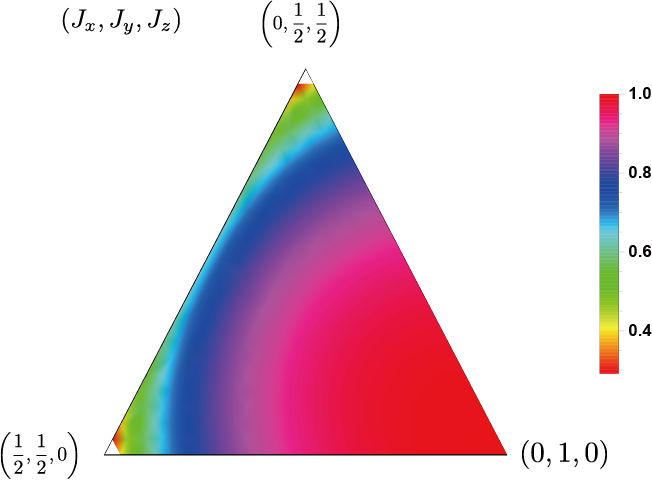}
\caption{
The $y$-component of the edge magnetization (per site)
 as a function of $(J_x,J_y,J_z)$ in A$_\mathrm{y}$ phase.
The system size used for the numerical calculation is $L_x =L_y  =32$.   
}
\label{Fig4}
\end{center}
\end{figure}

In Fig.~\ref{Fig4}, we plot the parameter dependence of the magnetization in A$_{\mathrm{y}}$ phase.  
It takes the maximum value $1$ for $(J_x,J_y,J_z) = (0,1,0)$,
since the edge spins completely behave as usual free spins in this limit.  
As the values of the parameters, $J_x, J_y, J_z$, approach the phase boundary, 
the magnetization decreases to a small value. 

Before concluding this section, we remark the following: 
For the edge magnetization in the $y$-direction which we have calculated in above, 
we have had to deal with the fermion parity operator $e^{i\pi \hat{\theta}_{(\ell,m)}}$, 
since $\sigma_{(1,m)}^y$ contains it. In Appendix~\ref{Sec:appB}, we present an alternative method  
to calculate the edge magnetization. More precisely, we introduce an alternative 
Jordan-Wigner transformation which is different from that used in Sec.~\ref{sec:model}. 
By using the transformation, we can obtain the exactly same result as Eq. (\ref{eq:edgemag_final}) 
under a certain assumption on the boundary condition. See Appendix~\ref{Sec:appB} for the details. 
We should also remark that a similar situation was already treated in Ref.~\onlinecite{Willans2011}, 
where the local magnetization induced by site vacancies was calculated.

\section{Discussions \label{sec:discussion}}
\subsection{Stability of edge magnetization}
We first address the stability of the edge magnetization against perturbations such as disorder and additional interactions. 

In the case of disorder, as we have shown in Sec. \ref{sec:edge}, the existence of the Majorana edge flat band is ensured by the weak topological nature of the Hamiltonian.
Namely, the Majorana edge flat band and the resulting edge magnetization are stable against perturbations which do not break the symmetries
of BDI class.  
For instance, consider the Kitaev honeycomb model with bond disorder whose Hamiltonian is given by
\begin{equation}
H = \sum_{\gamma\in\{x,y,z\}}\; \sum_{\langle i,j\rangle\in {\cal B}_\gamma} J_\gamma [i,j] \sigma_i^\gamma
\sigma_j^\gamma. 
\label{eq:ham_disorder}
\end{equation}
When the bond-dependent exchange integrals $J_\gamma[i,j]$ satisfy the conditions 
\begin{eqnarray}
|J_x [i,j]| +|J_z [i,j^{\prime \prime}]| < \kappa |J_y[i,j^\prime]|,
\end{eqnarray}
with $\langle i,j \rangle\in {\cal B}_x$, $\langle i,j^\prime \rangle\in {\cal B}_y$, and $\langle i,j^{\prime \prime} \rangle\in {\cal B}_z$,
and $\kappa \in (0,1)$, 
at every site $i$ on a honeycomb lattice,
we can construct the edge zero mode (for details, see Appendix \ref{sec:appC}). 

When a perturbation is an interaction between fermions which cannot be 
mapped to a free-fermion form,
the Hamiltonian cannot be diagonalized in terms of Majorana fermions in 
general.
Actually, if we introduce a standard Heisenberg interaction into the 
Kitaev spin Hamiltonian,
then a pair of $d$-Majorana fermions on $z$-bonds is no longer conserved.
It is a highly nontrivial problem whether or not the edge zero modes 
still survive under such a perturbation.
Since the free-fermion picture does not work well in this situation, we 
recall the valence bond picture, or,
the idea of paired and unpaired fermions. Consider the situation that 
all the couplings for the $y$ bonds are
very strong compared to the other couplings.
Then, we can expect that the internal degrees of freedom for all the 
pairs of two spins connected by the $y$ bonds are frozen,
and that there remains a possibility that only the spins at the edges 
are not frozen and behave as a free spin.
However, clearly, if we introduce a direct interaction between the spins 
at the edges,
then we cannot expect that the edge zero modes survive.
What is the necessary condition that the edge zero modes survive?
As we showed in Sec.~\ref{sec:model}, in the BdG representation (\ref{eq:BDG}),
the present unperturbed Kitaev Hamiltonian has 
time-reversal and particle-hole symmetries.
Roughly speaking, these symmetries protect the edge flat band against 
perturbations.
If the excitations of the quasi-particles near zero energy effectively 
preserve these symmetries under the additional interactions,
then we can expect that the edge flat band is not destroyed,
as long as
the bulk energy gap does not collapse, i.e., the system remains in a 
gapped spin liquid phase.

\subsection{Comparison with related works}
We briefly comment on related works. 
In the Kitaev model with site vacancy, 
localized zero energy states were found near the vacancy sites~\cite{Willans2011,Sreejith2016},
and these states yield a similar local magnetization to those in the present paper. 
For the $SU(2)$-symmetric version of the Kitaev model~\cite{Carvalho2018}, 
an edge magnetization was obtained recently.  
This edge magnetization also comes from the localized zero energy states. 
Thus, the appearance of the local magnetization is ubiquitous in the Kitaev-type models.  
Needless to say, in many models, various kinds of topological boundary states 
such as chiral edge modes~\cite{Yao2007} 
and corner modes~\cite{Dwivedi2018} have been found to appear. 

In the present paper, we have clarified the topological origin of the Majorana 
edge flat band which yields the edge magnetization by relying on the bulk-edge correspondence 
from the weak topological point of view. The advantage of this approach is that 
the stability of the Majorana edge flat band is guaranteed by the topological nature.
 This is very important from an experimental point of view because an additional perturbation 
such as disorder is often inevitable in experiments. 
However, we cannot explain the topological origin for all of the similar localized modes as in above 
by using our approach. Elucidating the topological origin of various boundary states 
is left for future studies. 

\subsection{Measurement of edge magnetization in experiments}
Finally, we address the possible experimental setup to measure the edge 
magnetization.
In the present Kitaev magnet, the highly anisotropic exchange integrals 
yield the large bulk gap above the ground state.
Our result implies that the situation also leads to the large edge 
magnetization.
Such a highly anisotropic magnet may be experimentally feasible 
by applying a pressure
to the Kitaev-magnet candidates~\cite{Bastien2018}.
But the additional interactions, such as the Heisenberg interaction, are also enhanced by the pressure.
Hence, it would be desirable to seek novel materials having the bond-anisotropic Kitaev interactions
to see the edge magnetization. 
If such a material is found, the highly unidirectional magnetization is expected to be detected
by applying a weak external magnetic field at one of the edges of the 
sample, as far as neither the Zeeman energy nor the thermal fluctuations exceed 
the flux gap in the gauge-field sector
and the bulk band gap of the itinerant Majorana fermions. We also stress 
the following:
As we have shown, this magnetization appears only at the edges because 
the bulk is in the gapped spin liquid phase.
Namely, if the external magnetic field is sufficiently weak, then the 
contribution of the magnetization from the bulk region
can be expected to be negligible.

\section{Summary \label{sec:summary}}
In summary, 
we have shown that the fractionalization of spins into Majorana fermions 
leads to a unidirectional edge magnetization in the Kitaev honeycomb model in a gapped phase
on a cylinder geometry.
The nonvanishing magnetization in a specific direction comes from the degeneracy of the ground state caused by the Majorana edge flat band
and the fermion parity conservation of localized Majorana fermions. 
Since the Majorana edge flat band is stable due to a weak topological nature of the BdG Hamiltonian,
the resulting edge magnetization is stable against the symmetry-preserving perturbations including disorders.
We hope that our results shed light on the new way to detect the Majorana fermions in the condensed matter systems.
  
\acknowledgements
We thank Y. Akagi, H. Katsura, M. Kohmoto, H. Tasaki, and M. Udagawa for the fruitful discussions. 
T. M. is supported by the JSPS KAKENHI (Grant No. JP17H06138), MEXT, Japan.

\appendix
\section{Bulk-edge correspondence in one dimension \label{sec:app}}
In this Appendix, we present a proof of the bulk-edge correspondence in one-dimensional 
tight-binding model for all the classes which show a topologically nontrivial index.
Here the bulk-edge correspondence means the equality of the bulk index (Ind$_\mathrm{B}$) and the edge index (Ind$_\mathrm{E}$), defined below. 

We begin our discussion with
AIII, BDI, and CII classes, where the system has chiral symmetry, and topological phases are characterized by the winding number.
For these classes, the proof has already been given by previous works~\cite{PSB,GS}. Here we give a proof that is slightly different from theirs. 
Note that the Kitaev honeycomb model, which belongs to BDI class,
indeed has zero-energy edge modes whose number is equal to the bulk winding number, as we have seen in the main text.

We also prove the bulk-edge correspondence for D and DIII classes which are characterized by a $\mathbb{Z}_2$ index.
As far as we know, the proof for these cases is new, although it is slightly similar to that for AIII, BDI, and CII classes having a $\mathbb{Z}$ index.

\subsection{AIII, BDI and CII Classes}
We first consider AIII, BDI and CII classes characterized by the winding number.
In the following, we denote by $O^\ast$ the adjoint of the operator $O$.

Let $S$ be the chiral operator which satisfies $S^2=1$ and $S^\ast=S$. 
Then, the chiral symmetric 
bulk Hamiltonian $H$ satisfies 
\begin{equation}
\label{chiralH}
SHS=-H. 
\end{equation} 
We assume that the hopping amplitudes of the tight-binding model are of finite range. 
Further, we require the following two assumptions: 
\begin{itemize}
\item The energy $E=0$ is not an eigenvalue of the Hamiltonian $H$. 
\item The energy $E=0$ is in the spectral gap of the Hamiltonian $H$ or in the localization regime. 
More precisely, we assume that the resolvent exponentially decays with distance as 
\begin{equation}
\sup_{\varepsilon>0}\Vert\chi_{x}(i\varepsilon-H)^{-1}\chi_{y} \Vert 
\le C_0\exp[-|x-y|/\xi_0],
\end{equation}
where $\chi_{x}$ is the characteristic function of the site $x$, and the two positive constants, 
$C_0$ and $\xi_0$, depend only on the parameters of the tight-binding model.  
\end{itemize}

We write $P_\pm$ for the spectral projections onto the positive and negative energies, respectively.  
We also write 
\begin{equation}
{\mathcal U}:=P_+-P_-
\end{equation}
for the flattened Hamiltonian,
and its eigenvalue is $+1$ ($-1$) if the corresponding eigenvalue of $H$ is positive (negative). 
Clearly, one has 
$S{\cal U}S=-{\cal U}$,
and 
\begin{equation}
{\cal U}=\left(
\begin{array}{cc}
0 & {\cal U}_- \\ 
{\cal U}_+ & 0 \\ 
\end{array}
\right),
\end{equation}
in the basis which diagonalizes the chiral operator $S$ as 
\begin{equation}
\label{diagonalS}
S=\left(
\begin{array}{cc}
1 & 0\\ 
0 & -1 \\
\end{array}
\right).
\end{equation}
One notices that the relation, ${\cal U}_-=({\cal U}_+)^\ast$, holds.  

Next, to discuss the bulk-edge correspondence, we need to
define the edge Hamiltonian $H_{\rm E}$ 
from the bulk Hamiltonian $H$.
To do this, we consider the half infinite chain with an open boundary 
at the left edge, 
and we introduce the projection operator $\cal{P}$ which restricts the whole infinite chain into the half one 
so that the edge Hamiltonian $H_{\rm E}$ is given by 
\begin{equation}
\label{edgeH}
H_{\rm E}={\cal{P}}H{\cal{P}}.
\end{equation}
Clearly, this projection operator $\cal{P}$ is equal to the step or switch function whose 
support is the half chain. Therefore, the commutator, $J:=i[H,{\cal P}]$, that is an operator 
on the whole chain, is the current operator across the left edge of the half chain. 
We assume that the chiral operator $S$ commutes with the projection operator ${\cal P}$, i.e., 
$[S,{\cal P}]=0$. When the chiral operator $S$ acts on the internal degrees of freedom at each site, 
this condition is obviously fulfilled. 

From (\ref{chiralH}) and (\ref{diagonalS}), the Hamiltonian $H$ can be written in the form  
\begin{equation}
H=\left( \begin{array}{cc}0 & H_- \\  H_+ & 0 \\ \end{array} \right), 
\end{equation}
with $H_-=(H_+)^\ast$. One may also write 
$
H=\tilde{H}_++\tilde{H}_-
$
with 
$\tilde{H}_\pm=H\cdot \frac{1}{2}(1\pm S).$
Then, the edge Hamiltonian $H_{\rm E}$ of (\ref{edgeH}) can be written as
\begin{equation}
H_{\rm E}={\cal P}\tilde{H}_+{\cal P}+{\cal P}\tilde{H}_+^\ast{\cal P}. 
\end{equation}
One notices that there are two types of zero modes. One is a set of eigenvectors of the chiral 
operator $S$ with eigenvalue $+1$. The other is that with the opposite eigenvalue $-1$. 
The integer-valued index of the edge zero modes is defined by 
\begin{eqnarray}
{\rm Ind}_{\rm E}&:=&{\rm Ind}({\cal P}H_+{\cal P}+1-{\cal P})\nonumber\\
&:=&{\rm dim}\;{\rm ker}\;({\cal P}H_+{\cal P}+1-{\cal P}) \nonumber\\
& -& {\rm dim}\;{\rm ker}\;({\cal P}H_+^\ast{\cal P}+1-{\cal P}), 
\end{eqnarray}
where ${\rm dim}\;{\rm ker}\;(O)$ stands for the dimension of the kernel of an operator $O$. 
Namely, the edge index, ${\rm Ind}_{\rm E}$, is defined by the difference between 
the numbers of the two types of the edge states. In general, when an operator $T$ satisfies 
${\rm dim}\;{\rm ker}\;(T)<\infty$ and ${\rm dim}\;{\rm ker}\;(T^\ast)<\infty$, 
the operator $T$ is called the Fredholm operator \cite{Palais,Bonic}, and the Fredholm index is defined by 
${\rm Ind}\;(T):={\rm dim}\;{\rm ker}\;(T)-{\rm dim}\;{\rm ker}\;(T^\ast)$.  
In the following, the edge index ${\rm Ind}_{\rm E}$ will prove to be the Fredholm index. 

In order to show that the edge index is equal to the bulk index, i.e., the winding number,  
we recall some results in Ref. \onlinecite{KatsuraKoma}. 
The bulk index is defined by 
\begin{equation}
\label{IndB}
{\rm Ind}_{\rm B}:=\frac{1}{2}{\rm Tr}\; [S({\cal P}-{\cal U}{\cal P}{\cal U})]. 
\end{equation}
By using the supersymmetric structure \cite{ASS} in the operator algebra, this index can be written as 
\begin{eqnarray}
{\rm Ind}_{\rm B}&=&{\rm Ind}({\cal P}{\cal U}_+{\cal P}+1-{\cal P})\nonumber\\
&=&{\rm dim}\;{\rm ker}\;({\cal P}{\cal U}_+{\cal P}+1-{\cal P})\nonumber\\
&-&{\rm dim}\;{\rm ker}\;({\cal P}{\cal U}_+^\ast{\cal P}+1-{\cal P}).
\end{eqnarray}
Thus, the bulk index ${\rm Ind}_{\rm B}$ takes an integer value. 

\begin{figure}[b]
\begin{center}
\includegraphics[width=0.95\linewidth]{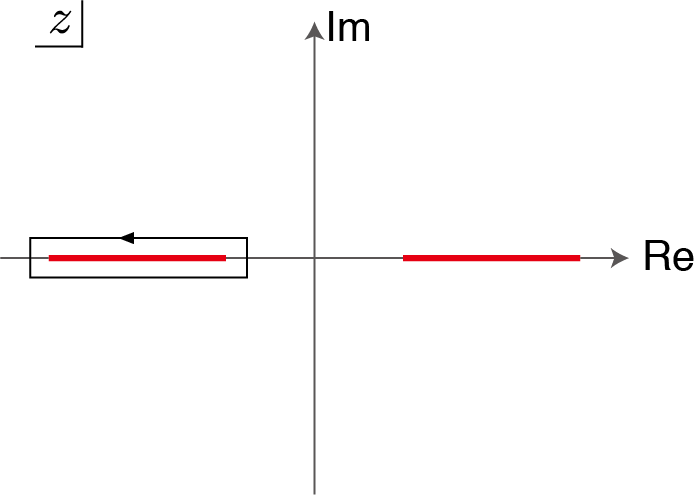}
\caption{
Schematic picture of the contour in Eq. (\ref{P-}).
Red lines denote the spectrum of $H$. 
}
\label{Fig5}
\end{center}
\end{figure}
The right-hand side of the bulk index (\ref{IndB}) can also be written in the form of the winding number as  
\begin{equation}
\label{windingnumber}
{\rm Ind}_{\rm B}=\frac{1}{2\pi}\oint dz\; {\rm Tr}\;S{\cal U}\frac{1}{z-H}J\frac{1}{z-H}. 
\end{equation}
In Eq. (\ref{windingnumber}), the contour integral in the complex plane 
is chosen so that the contour integral of the resolvent $(z-H)^{-1}$ yields 
the projection $P_-$ onto the sector of the negative energies of the Hamiltonian (see Fig.~\ref{Fig5}), i.e.,  
\begin{equation}
\label{P-}
P_-=\frac{1}{2\pi i}\oint \frac{dz}{z-H}.  
\end{equation}
The projection ${\cal P}$ onto the half chain is equal to the characteristic function of the half chain. 
Combining this with the assumption that the hopping amplitudes of the tight-binding model are of finite range, 
one notices that the current operator $J$ has a finite support. 
Besides, from the assumption that the resolvent $(z-H)^{-1}$ has the upper bound 
that exponentially decays with distance at the Fermi level 
$E=0$, the operator $(z-H)^{-1}J(z-H)^{-1}$ is trace class. 
Thus, the right-hand side of (\ref{windingnumber}) is well defined. 

The proof of the equality 
of the bulk index (\ref{IndB}) and the winding number (\ref{windingnumber}) is as follows:    
Note that 
\begin{equation}
{\cal P}-{\cal U}{\cal P}{\cal U}={\cal U}[{\cal U},{\cal P}]=-2{\cal U}[P_-,{\cal P}], \label{eq:a14}
\end{equation}
where we have used ${\cal U}^2=1$ and ${\cal U}=1-2P_-$. 
The commutator in the right-hand side of (\ref{eq:a14}) is written 
\begin{eqnarray}
[P_-,{\cal P}]&=&\frac{1}{2\pi i}\oint dz \frac{1}{z-H}[H,{\cal P}]\frac{1}{z-H} \nonumber \\
&=&-\frac{1}{2\pi}\oint dz \frac{1}{z-H}J\frac{1}{z-H},
\end{eqnarray}
where we have used the expression (\ref{P-}) of the projection $P_-$, and 
$J=i[H,{\cal P}]$. 
Substituting these into the right-hand side of (\ref{IndB}), 
one obtains the desired result (\ref{windingnumber}). 

Now, we give a proof of the bulk-edge correspondence, i.e., ${\rm Ind}_{\rm E}={\rm Ind}_{\rm B}$. 
The operator $H_+$ is the restriction of the Hamiltonian $H$ onto the sector  
$\mathcal{H}_+$ with eigenvalue $+1$ for the chiral operator $S$, i.e.,  
\begin{equation}
H_+=H\cdot\left.\frac{1}{2}(1+S)\right|_{\mathcal{H}_+}.
\end{equation}
Note that $H=|H|\mathcal{U}$, where $|H|=\sqrt{H^2}$. Using this polar decomposition, 
one has 
\begin{eqnarray}
{\cal P}H_+{\cal P}&=&\left.{\cal P}|H|{\cal U}{\cal P}\right|_{\mathcal{H}_+}\nonumber\\
&=&\left.{\cal P}|H|{\cal P}{\cal U}{\cal P}\right|_{\mathcal{H}_+}
+\left.{\cal P}|H|(1-{\cal P}){\cal U}{\cal P}\right|_{\mathcal{H}_+}\nonumber\\
&=&\left.{\cal P}|H|\right|_{\mathcal{H}_-}{\cal P}{\cal U}_+{\cal P}
+\left.{\cal P}|H|(1-{\cal P}){\cal U}{\cal P}\right|_{\mathcal{H}_+}. \nonumber\\
\label{H+decompo}
\end{eqnarray}
In order to treat the first term in the right-hand side of (\ref{H+decompo}), we recall the well known fact about 
Fredholm indices. (See, e.g., Refs. \onlinecite{Palais,Bonic}.) Let $T_1$ and $T_2$ be two Fredholm operators for which the product $T_1T_2$ 
is defined. Then, the product $T_1T_2$ is also the Fredholm operator, and the index satisfies the relation,  
\begin{equation}
\label{FredholmT1T2}
{\rm Ind}\;(T_1T_2)={\rm Ind}\;(T_1)+{\rm Ind}\;(T_2).
\end{equation}

Note that 
\begin{eqnarray}
\label{productTwoFredholm}
& (\left.{\cal P}|H|{\cal P}\right|_{\mathcal{H}_-}+1-{\cal P})({\cal P}{\cal U}_+{\cal P}+1-{\cal P}) \nonumber \\
&= \left.{\cal P}|H|\right|_{\mathcal{H}_-}{\cal P}{\cal U}_+{\cal P}
+1-{\cal P}.
\end{eqnarray}
From the assumption of the spectrum of the Hamiltonian $H$, one has $|H|>0$. Therefore, one obtains 
\begin{equation}
{\rm Ind}\;(\left.{\cal P}|H|{\cal P}\right|_{\mathcal{H}_-}+1-{\cal P})=0.
\end{equation}
By using this and the relation (\ref{FredholmT1T2}), we have 
\begin{eqnarray}
& &{\rm Ind}\;((\left.{\cal P}|H|{\cal P}\right|_{\mathcal{H}_-}+1-{\cal P})({\cal P}{\cal U}_+{\cal P}+1-{\cal P})) \nonumber \\
&=&{\rm Ind}\;(\left.{\cal P}|H|{\cal P}\right|_{\mathcal{H}_-}+1-{\cal P})
+{\rm Ind}\;({\cal P}{\cal U}_+{\cal P}+1-{\cal P}) \nonumber\\
&=&{\rm Ind}\;({\cal P}{\cal U}_+{\cal P}+1-{\cal P}).
\end{eqnarray}
Combining this with the identity (\ref{productTwoFredholm}), we obtain 
\begin{equation}
\label{Indexrelation}
{\rm Ind}\;(\left.{\cal P}|H|\right|_{\mathcal{H}_-}{\cal P}{\cal U}_+{\cal P}
+1-{\cal P})={\rm Ind}\;({\cal P}{\cal U}_+{\cal P}+1-{\cal P}). 
\end{equation}

Next, consider the second term in the right-hand side of (\ref{H+decompo}). As is well known, 
Fredholm indices are stable against a compact perturbation. Namely, for a Fredholm operator $T$ and 
a compact operator $K$, the following relation is valid: \cite{Palais,Bonic}
\begin{equation}
\label{compactperturb}
{\rm Ind}\;(T+K)={\rm Ind}\;(T).
\end{equation}
Therefore, it is sufficient to prove that the second term in the right-hand side of (\ref{H+decompo}) is 
compact. Actually if so, one can obtain the desired result, 
\begin{equation}
{\rm Ind}_{\rm E}={\rm Ind}\;({\cal P}H_+{\cal P}+1-{\cal P})=
{\rm Ind}\;({\cal P}{\cal U}_+{\cal P}+1-{\cal P})={\rm Ind}_{\rm B},
\end{equation}
from (\ref{H+decompo}), (\ref{Indexrelation}) and (\ref{compactperturb}). 

Using ${\cal U}=1-2P_-$ and the expression (\ref{P-}) of the projection $P_-$, we have 
\begin{eqnarray}
(1-{\cal P}){\cal U}{\cal P}&=&(1-{\cal P})(1-2P_-){\cal P}  \nonumber \\
&=&-2\times\frac{1}{2\pi i}\oint dz\; (1-{\cal P})\frac{1}{z-H}{\cal P}.\nonumber \\ \label{eq:a25}
\end{eqnarray}
From the assumption that the resolvent $(z-H)^{-1}$ exponentially decays with distance at 
the Fermi level $E=0$, this right-hand side of (\ref{eq:a25}) is compact. 
Thus, the second term in the right-hand side of (\ref{H+decompo}) is compact because $|H|$ is bounded. 

\subsection{D class}
In the following two subsections, we treat the classes which show a nontrivial $\mathbb{Z}_2$ index.
For D class, the Hamiltonian $H$ has only a particle-hole symmetry.  
Namely, for an anti-linear transformation $\Xi$ and for any wavefunction $\varphi$, 
the following relation holds:
\begin{equation}
\Xi H\varphi=-H\Xi\varphi. 
\end{equation}
Similarly to the preceding case, we write $P_\pm$ for the spectral projection onto the positive and negative energies of the 
Hamiltonian $H$, respectively. We also write ${\cal U}:=P_+-P_-$ for the flattened Hamiltonian.  
Clearly, one has 
\begin{equation}
\Xi{\cal U}\varphi=-{\cal U}\Xi\varphi
\end{equation}
for any wavefunction $\varphi$. 
In the present case, we assume that the Fermi level $E=0$ lies in a nonvanishing spectral gap of the Hamiltonian $H$. 
For the case of the mobility gap, our approach below does not work well. This is left for future studies. 

The edge Hamiltonian $H_{\rm E}$ is defined by $H_{\rm E}:={\cal P}H{\cal P}$, 
where ${\cal P}$ is the restriction of the whole infinite chain to the half infinite chain. 
We also assume 
\begin{equation}
\Xi{\cal P}\varphi={\cal P}\Xi\varphi
\end{equation}
for any wavefunction $\varphi$. Namely, the particle-hole transformation $\Xi$ acts on only 
the internal degree of freedom at each site. 

In this class, the bulk $\ze_2$ index is defined by~\cite{KatsuraKoma}
\begin{equation}
{\rm Ind}_{\rm B}^{(2)}:={\rm dim}\;{\rm ker}\;({\cal P}{\cal U}{\cal P}+1-{\cal P})\ \ {\rm modulo}\ 2,
\end{equation}
and the edge $\ze_2$ index is defined by 
\begin{equation}
{\rm Ind}_{\rm E}^{(2)}:={\rm dim}\;{\rm ker}\;(H_{\rm E})\ \ {\rm modulo}\ 2. 
\end{equation}
Clearly, this is equal to the even-oddness of the number of the edge zero modes. 

Let us give a proof of the equality of the bulk and edge indices. To begin with, we note that 
\begin{eqnarray}
H={\cal P}H{\cal P}+{\cal P}H(1-{\cal P})+(1-{\cal P})H{\cal P}+(1-{\cal P})H(1-{\cal P}). \nonumber \\ 
\end{eqnarray}
Since the range of the hopping amplitudes of the Hamiltonian $H$ is finite, 
the second and third terms in the right-hand side are compact operators. 
Therefore, the essential spectrum of $H$ is equal to the essential spectrum of 
${\cal P}H{\cal P}+(1-{\cal P})H(1-{\cal P})$. (For the stability of an essential spectrum 
under a compact perturbation, see, e.g, Ref. \onlinecite{Kato}.) This implies that the spectrum of the edge Hamiltonian 
$H_{\rm E}={\cal P}H{\cal P}$ 
which is restricted to the region of the spectral gap of $H$ is only a subset of the discrete spectrum of 
$H_{\rm E}$. 

The edge Hamiltonian $H_{\rm E}$ can be written as 
\begin{eqnarray}
H_{\rm E}={\cal P}H{\cal P}={\cal P}|H|{\cal U}{\cal P}
={\cal P}|H|{\cal P}{\cal U}{\cal P}+{\cal P}|H|(1-{\cal P}){\cal U}{\cal P}. \nonumber\\ \label{eq:classd_edgeham}
\end{eqnarray}
From the assumption of the spectral gap of the Hamiltonian $H$, in the same way as in the preceding case, 
one can prove that the second term in the right-hand side of (\ref{eq:classd_edgeham}) is a compact operator. 
We introduce a Hamiltonian, 
\begin{equation}
H_{\rm E}(g):=  {\cal P}|H|{\cal P}{\cal U}{\cal P}+g{\cal P}|H|(1-{\cal P}){\cal U}{\cal P}, 
\end{equation}
with the parameter $g\in[0,1]$. Clearly, $H_{\rm E}(1)=H_{\rm E}$. 
Let $\varphi$ be an eigenvector of $H_{\rm E}(g)$ with eigenvalue $\lambda\ne 0$, i.e., 
$H_{\rm E}(g)\varphi=\lambda\varphi$.
Then, one has 
\begin{equation}
H_{\rm E}(g)\Xi\varphi=-\lambda\Xi\varphi.
\end{equation}
Thus, $\lambda$ and $-\lambda$ come in pairs of eigenvalues of $H_{\rm E}(g)$. 
On the other hand,  the discrete spectrum of $H_{\rm E}(g)$ is continuous with respect to the compact perturbation 
in the spectral gap region. From these observations, we conclude that the even-oddness of the number of the zero modes of 
$H_{\rm E}(g)$ does not depend on the parameter $g$, i.e., 
\begin{eqnarray}
{\rm dim}\;{\rm ker}\;(H_{\rm E}) = & {\rm dim}\;{\rm ker}\;(H_{\rm E}(1)) \nonumber \\
= & {\rm dim}\;{\rm ker}\;(H_{\rm E}(0))\ \ {\rm modulo}\ 2. \nonumber \\
\end{eqnarray}
The right-hand side can be written as   
\begin{eqnarray}
{\rm dim}\;{\rm ker}\;(H_{\rm E}(0))&=&{\rm dim}\;{\rm ker}\;({\cal P}|H|{\cal P}{\cal U}{\cal P}+1-{\cal P}) \nonumber \\
&=&{\rm dim}\;{\rm ker}\;({\cal P}{\cal U}{\cal P}+1-{\cal P}),
\end{eqnarray}
where we have used $|H|>0$ which can be derived from the assumption of the spectral gap at the Fermi level $E=0$. 
These imply the bulk-edge correspondence, ${\rm Ind}_{\rm E}^{(2)}={\rm Ind}_{\rm B}^{(2)}$, by the definitions.  

\subsection{DIII class}
In the final case, the Hamiltonian $H$ has three symmetries, time-reversal, particle-hole and chiral symmetries 
whose transformations are, respectively, denoted by $\Theta$, $\Xi$ and $S$. 
Then, the bulk Hamiltonian $H$ is transformed as 
\begin{equation}
\Theta H\varphi=H\Theta\varphi, \quad \Xi H\varphi=-H\Xi\varphi \quad \mbox{and} \quad 
SH\varphi=-HS\varphi
\end{equation}
for any wavefunction $\varphi$. The two anti-linear transformations, $\Theta$ and $\Xi$, satisfy 
\begin{equation}
\Theta^2\varphi=-\varphi \quad \mbox{and} \quad \Xi^2\varphi=\varphi.
\end{equation}
Further, the relation, $\Xi=S\Theta$, holds. We assume that the Fermi level $E=0$ lies in a nonvanishing 
spectral gap of the Hamiltonian $H$.  

Similarly to the previous two cases, we define the edge Hamiltonian by $H_{\rm E}:={\cal P}H{\cal P}$ 
with the restriction ${\cal P}$ of the whole infinite chain to the half infinite chain. 
We assume that each of the three transformations, $\Theta$, $\Xi$ and $S$, commutes with the restriction ${\cal P}$. 
Since the Hamiltonian $H$ has the chiral symmetry, the edge Hamiltonian $H_{\rm E}$ is decomposed into 
two parts, 
\begin{equation}
H_{\rm E}={\cal P}\tilde{H}_+{\cal P}+{\cal P}\tilde{H}_-{\cal P}
\end{equation}
in the same way as in the first case. Therefore, there are two types of zero mode which are also 
the eigenvectors of the chiral operator $S$ with eigenvalue $+1$ and those with eigenvalue $-1$. 

Let $\varphi$ be a zero edge mode, i.e., $H_{\rm E}\varphi=0$. Then, from the time-reversal symmetry of 
the Hamiltonian $H$ and the assumption that the time-reversal transformation $\Theta$ commutes with ${\cal P}$, 
$\Theta\varphi$ is also a zero mode. These two states form the Kramers doublet because of the odd time-reversal 
symmetry $\Theta^2=-1$. 
Namely, the two states satisfy $\braket{\varphi | \Theta\varphi}=0$. 
In addition to $H_{\rm E}\varphi=0$, let $\varphi$ be an eigenvector of the chiral operator $S$ with eigenvalue $+1$, i.e., 
$S\varphi=\varphi$. Then, one has $S\Theta\varphi=-\Theta\varphi$.  
In order to prove this statement, we note that 
\begin{equation}
S\Theta S\Theta\psi=\psi
\end{equation}
for any wavefunction $\psi$. This can be obtained from $\Xi=S\Theta$ and $\Xi^2=+1$. Further, by using 
$S^2=1$ and $\Theta^2=-1$, one has 
\begin{equation}
S\Theta\psi=-\Theta S\psi. 
\end{equation}
This yields $S\Theta\varphi=-\Theta\varphi$ for the wavefunction $\varphi$ satisfying $S\varphi=\varphi$.
Thus, the two types of the zero modes have the same degeneracy. This implies that the degeneracy of the edge zero modes 
is always even. 
Relying on this fact, we define the edge $\ze_2$ index by 
\begin{eqnarray}
\label{IndE2}
{\rm Ind}_{\rm E}^{(2)}:= & \frac{1}{2}{\rm dim}\;{\rm ker}\;(H_{\rm E}) \nonumber \\ 
= & \frac{1}{2}{\rm dim}\;{\rm ker}\;({\cal P}\tilde{H}_+{\cal P}+{\cal P}\tilde{H}_-{\cal P})\ \ \mbox{modulo}\ 2. \nonumber \\
\end{eqnarray}

In the following, we will use the same notations, $P_\pm$, ${\cal U}$, and ${\cal U}_\pm$, as in the first case 
because the Hamiltonian $H$ has the chiral symmetry in the present case. 
The bulk $\ze_2$ index is defined by 
\begin{equation}
\label{IndB2}
{\rm Ind}_{\rm B}^{(2)}:={\rm dim}\;{\rm ker}\;({\cal P}{\cal U}_+{\cal P}+1-{\cal P})\quad \mbox{modulo}\ 2.
\end{equation}
 
By using the polar decomposition $H=|H|{\cal U}$,  the edge Hamiltonian $H_{\rm E}$ can be written as 
\begin{equation}
H_{\rm E}={\cal P}|H|{\cal U}{\cal P}={\cal P}|H|{\cal P}{\cal U}{\cal P}
+{\cal P}|H|(1-{\cal P}){\cal U}{\cal P}.
\end{equation}
We again introduce
\begin{equation}
\label{HE(g)}
H_{\rm E}(g):={\cal P}|H|{\cal P}{\cal U}{\cal P}+g{\cal P}|H|(1-{\cal P}){\cal U}{\cal P}
\end{equation}
with the parameter $g\in[0,1]$. Clearly, one has $H_{\rm E}=H_{\rm E}(1)$. Note that 
\begin{equation}
SH_{\rm E}(g)\varphi=-H_{\rm E}(g)S\varphi
\end{equation}
and
\begin{equation}
\Xi H_{\rm E}(g)\varphi=-H_{\rm E}(g)\Xi\varphi
\end{equation}
for any wavefunction $\varphi$. These imply that, if $\varphi$ is an energy eigenvector of 
the edge Hamiltonian $H_{\rm E}$ with eigenvalue $\lambda$, $S\varphi$ and $\Xi\varphi$ are also 
an energy eigenvector of $H_{\rm E}$ with eigenvalue $-\lambda$.   
Further, we have  
\begin{eqnarray}
\braket{ S \varphi | \Xi \varphi}   =  & \braket{ S \varphi | S\Theta \varphi} \nonumber \\
 =  & \braket{  \varphi |  \Theta \varphi} =0, 
\end{eqnarray}
where we have used $\Xi=S\Theta$, $S^\ast S=1$ and $\Theta^2=-1$. Thus, $\lambda$ and $-\lambda$ come in pairs 
of the eigenvalues of $H_{\rm E}(g)$ with opposite sign, and both of them have even degeneracy. 

Combining these observations about the discrete spectrum of $H_{\rm E}(g)$ 
with the fact that the second term in the right-hand side of (\ref{HE(g)}) is compact, we have  
\begin{eqnarray}
\label{dimkerHE}
\frac{1}{2}{\rm dim}\;{\rm ker}\;(H_{\rm E}) 
= & \frac{1}{2}{\rm dim}\;{\rm ker}\;(H_{\rm E}(1)) \nonumber \\
= & \frac{1}{2}{\rm dim}\;{\rm ker}\;(H_{\rm E}(0))
\quad \mbox{modulo}\ 2. \nonumber \\
\end{eqnarray}

On the other hand, we have 
\begin{widetext}
\begin{eqnarray}
{\rm ker}\;(H_{\rm E}(0))&=&{\rm ker}\;({\cal P}|H|{\cal P}{\cal U}{\cal P}+1-{\cal P}) \nonumber\\
&=&{\rm ker}\;\left({\cal P}|H|{\cal P}{\cal U}{\cal P}\cdot\frac{1}{2}(1+S)
+{\cal P}|H|{\cal P}{\cal U}{\cal P}\cdot\frac{1}{2}(1-S)+1-{\cal P}\right) \nonumber \\
&=&{\rm ker}\;\left(\left.{\cal P}|H|{\cal P}{\cal U}_+{\cal P}\right|_{{\cal H}_+}
+\left.{\cal P}|H|{\cal P}{\cal U}_-{\cal P}\right|_{{\cal H}_-}+1-{\cal P}\right) \nonumber\\
&=&{\rm ker}\;(\left.{\cal P}|H|{\cal P}{\cal U}_+{\cal P}+1-{\cal P})\right|_{{\cal H}_+}
+{\rm ker}\;(\left.{\cal P}|H|{\cal P}{\cal U}_-{\cal P}+1-{\cal P})\right|_{{\cal H}_-} \nonumber \\
&=&{\rm ker}\;(\left.{\cal P}{\cal U}_+{\cal P}+1-{\cal P})\right|_{{\cal H}_+}
+{\rm ker}\;(\left.{\cal P}{\cal U}_-{\cal P}+1-{\cal P})\right|_{{\cal H}_-},
\end{eqnarray}
\end{widetext}
where we have used $|H|>0$ that is obtained from the assumption of the spectral gap of the Hamiltonian $H$. 
Therefore, we obtain 
\begin{widetext}
\begin{eqnarray}
{\rm dim}\;{\rm ker}\;(H_{\rm E}(0))
&=&{\rm dim}\;{\rm ker}\;(\left.{\cal P}{\cal U}_+{\cal P}+1-{\cal P})\right|_{{\cal H}_+}
+{\rm dim}\;{\rm ker}\;(\left.{\cal P}{\cal U}_-{\cal P}+1-{\cal P})\right|_{{\cal H}_-} \nonumber \\
&=&2\;{\rm dim}\;{\rm ker}\;(\left.{\cal P}{\cal U}_+{\cal P}+1-{\cal P})\right|_{{\cal H}_+}, 
\end{eqnarray}
\end{widetext}
where we have used the relation,
\begin{eqnarray}
 & {\rm dim}\;{\rm ker}\;(\left.{\cal P}{\cal U}_+{\cal P}+1-{\cal P})\right|_{{\cal H}_+} \nonumber \\ 
& =  {\rm dim}\;{\rm ker}\;(\left.{\cal P}{\cal U}_-{\cal P}+1-{\cal P})\right|_{{\cal H}_-},
\end{eqnarray}
that holds in the present case \cite{KatsuraKoma}. Combining this, (\ref{IndE2}), (\ref{IndB2}) and (\ref{dimkerHE}), 
we obtain the desired result, ${\rm Ind}_{\rm E}^{(2)}={\rm Ind}_{\rm B}^{(2)}$. 

\section{An alternative method to calculate the edge magnetization \label{Sec:appB}}
In the calculation of the edge magnetization discussed in the main text, 
the difficulty comes from the treatment of the fermion parity operator $e^{i \pi \hat{\theta}_{(\ell,m)}}$.
In this appendix, we explain an alternative approach to calculate the edge magnetization, by which we can avoid the difficulty.
This approach is similar to one discussed in Ref.~\cite{Willans2011} where  
the Hamiltonian with the Zeeman term for the external magnetic field 
has a quadratic form of the Majorana fermion. The approach in Ref.~\cite{Willans2011} 
employs the Kitaev's Majorana representation, while we use an alternative Jordan-Wigner 
transformation which is different from that in Sec.~\ref{sec:model}. 
Although our approach has an advantage that it is free from the projection procedure~\cite{Willans2011}
onto the physical space after using the Kitaev's Majorana representation, 
we must change the boundary condition in the vertical direction from open to periodic. 

We first introduce a unitary transformation, $ (\sigma^x, \sigma^y,\sigma^z) \rightarrow (\sigma^y, \sigma^z,\sigma^x)$,  
which is rotation by $\frac{2\pi}{3}$ about the axis in the (1,1,1) direction. 
Then, the Hamiltonian of Eq. (\ref{eq:hamspin}) is transformed as 
\begin{eqnarray}
H = \sum_{\langle i,j\rangle\in {\cal B}_x} J_x \sigma_i^y \sigma_j^y 
+ \sum_{\langle i,j\rangle\in {\cal B}_y} J_y \sigma_i^z \sigma_j^z 
+ \sum_{\langle i,j\rangle\in {\cal B}_z} J_z \sigma_i^x \sigma_j^x. \nonumber \\
\label{eq:hamspin2}
\end{eqnarray}

Next, we perform the Jordan-Wigner transformation in a different manner from Eqs. (\ref{eq:JW_sigmax})-(\ref{eq:JW_sigmaz}). 
This transformation is useful when an open boundary condition is imposed in both horizontal and vertical directions,
because the fermion parity operator does not appear in the fermion Hamiltonian.  
To do this, 
we first introduce a set of sites in $2j-1$-th and $2j$-th columns, as $X_j =\{ (2j-1,1),(2j,1),(2j-1,2),(2j,2), \cdots (2j-1,L_y) , (2j,L_y) \}$ with $j = 1, \cdots, L_x$.
In order to construct the fermion parity operator $e^{i\pi \hat{\phi}_{(\ell,m)}}$ in (\ref{eq:JW_sigmax2}) and (\ref{eq:JW_sigmay2}) below, 
we further introduce an order relation $\prec$ to the site set $X_j$ so that $X_j$ becomes a totally ordered set. 
The order relation is defined as:  
$$
(2j-1, m^\prime) \prec (2j-1, m) \hspace{2mm} \mathrm{if} \hspace{2mm} m^\prime < m, 
$$
$$
(2j, m^\prime) \prec (2j, m) \hspace{2mm} \mathrm{if} \hspace{2mm} m^\prime < m,
$$
$$
(2j-1, m^\prime) \prec (2j, m) \hspace{2mm} \mathrm{if} \hspace{2mm} m^\prime \leq m,
$$
$$
(2j, m^\prime) \prec (2j-1, m) \hspace{2mm}  
\hspace{8mm} \mathrm{if} \hspace{2mm} m^\prime < m \hspace{2mm} 
$$
and 
$$
(\ell,m) \nprec  (\ell,m) \hspace{2mm} \mathrm{for} \hspace{2mm} (\ell,m) \in X_j.
$$

Then, the Jordan-Wigner transformation is given by
\begin{equation}
\sigma_{(\ell,m)}^+ = 2 a_{(\ell,m)}e^{i\pi\hat{\phi}_{(\ell,m)}}, \label{eq:JW_sigmax2}
\end{equation}
\begin{equation}
\sigma_{(\ell,m)}^- = 2e^{i\pi\hat{\phi}_{(\ell,m)}} a^\dagger_{(\ell,m)}, \label{eq:JW_sigmay2}
\end{equation}
\begin{equation}
\sigma_{(\ell,m)}^z = (-1)^\ell \left[ 2a_{(\ell,m)}^{\dagger}a_{(\ell,m)} -1 \right], \label{eq:JW_sigmaz2}
\end{equation}
with
\begin{eqnarray}
\hat{\phi}_{(\ell,m)} &=&\sum_{j =1}^{[\frac{\ell + 1}{2}] -1}
\sum_{(\ell^\prime , m^\prime) \in X_j} 
a^{\dagger}_{(\ell^\prime,m^\prime)}a_{(\ell^\prime,m^\prime)} \nonumber \\
&+& \sum_{(\ell^\prime , m^\prime) \in X_{[\frac{\ell + 1}{2}]}: (\ell^\prime , m^\prime) \prec (\ell,m)} a^{\dagger}_{(\ell^\prime,m^\prime)}a_{(\ell^\prime,m^\prime)} ,
\nonumber \\
\end{eqnarray}
where $[ \cdots ]$ is Gauss symbol.
We further introduce Majorana fermions as 
\begin{eqnarray}
c_{(\ell,m)} = &a_{(\ell,m)}^\dagger + a_{(\ell,m)}, \nonumber\\
d_{(\ell,m)} =  & i\left[a_{(\ell,m)}^\dagger - a_{(\ell,m)}\right]\ \ \mbox{if $\ell$ is odd}, \label{eq:Majoranaodd2}
\end{eqnarray}
and\begin{eqnarray}
c_{(\ell,m)} = &  i \left[a_{(\ell,m)}^\dagger -a_{(\ell,m)}\right], \nonumber\\
d_{(\ell,m)} = & a_{(\ell,m)}^\dagger + a_{(\ell,m)}\ \ \mbox{if $\ell$ is even}. \label{eq:Majoranaeven2}
\end{eqnarray} 
Then, the Hamiltonian can be written as 
\begin{eqnarray}
H &=&  i J_x \sum_{\ell = 1}^{L_x} \sum_{m = 1}^{L_y} c_{(2\ell-1 ,m)} c_{(2\ell,m )} \nonumber\\
 &+&  J_y\sum_{\ell = 1}^{L_x-1} \sum_{m = 1}^{L_y} c_{(2\ell ,m)} c_{(2\ell+1 ,m)} d_{(2\ell ,m)} d_{(2\ell+1 ,m)} \nonumber\\
&+& i J_z \sum_{\ell = 1}^{L_x } \sum_{m = 1}^{L_y-1} c_{(2\ell,m)} c_{(2\ell- 1,m+1)}.  \nonumber\\
\label{eq:Hamiltonian_Majorana2}
\end{eqnarray}
Clearly, the conserved quantities, $d_{(2\ell,m)} d_{(2\ell +1,m)}$, live on the $y$-bonds.
Similarly to the case in Sec.~\ref{sec:model}, we choose $d_{(2\ell,m)} d_{(2\ell +1,m)} = i$ for all $\ell = 1, \cdots, L_x-1$ and $m = 1, \cdots, L_y$.
Then, the Hamiltonian (\ref{eq:Hamiltonian_Majorana2}) is exactly equal to the Hamiltonian (\ref{eq:hami}) in Sec.~\ref{sec:model} except for the boundary conditions. 

Now, let us consider the case where the external magnetic field in the $z$-direction is applied only at the left edge.
The Hamiltonian of the Zeeman energy is written as 
\begin{equation}
H_{\mathrm{Z}} =  - i h\sum_{m=1}^{L_y}  d_{(1,m)} c_{(1,m)} \label{eq:ham_zeeman}
\end{equation}
with the real parameter $h$. 
Then, one notices that 
the total Hamiltonian,
$H + H_{\mathrm{Z}}$, can be written as a quadratic form of 
$c_{(\ell,m)}$ and $d_{(1,m)}$. 
In fact, since $d_{(1,m)}$ is not included in the Hamiltonian of Eq. (\ref{eq:Hamiltonian_Majorana2}),
the ground state in the presence of the magnetic field, 
$\ket{\mathrm{GS}(h)}$, can be obtained, in principle, by diagonalizing the Hamiltonian which is written in the quadratic form of the Majorana fermions.
Then the edge magnetization for the $z$-component of spin is given by
\begin{eqnarray}
&& \frac{1}{L_y}\sum_{m=1}^{L_y} \bra{\mathrm{GS}(h)} \sigma^z_{(1,m)}\ket{\mathrm{GS}(h)}  \nonumber \\
&=& \frac{1}{L_y} \sum_{m=1}^{L_y} \bra{\mathrm{GS}(h)} 
i d_{(1,m)} c_{(1,m)} \ket{\mathrm{GS}(h)}.
\end{eqnarray}

In order to obtain the quadratic form of the Majorana fermions for the Hamiltonian, we have 
imposed the open boundary condition in the vertical direction in above.  
But it is fairly difficult to calculate the edge magnetization for the open boundary condition. 
Therefore, we change the boundary condition in the vertical direction from open to periodic 
for the Hamiltonian with the quadratic form of the Majorana fermions. 
When the side length $L_y$  of the lattice in the vertical direction is sufficiently large, 
we can expect that the effect of boundary conditions does not affect the value of 
the edge magnetization per length of the edge.  Clearly, when we impose the periodic 
boundary condition, the system is translationally invariant in the vertical direction. 
Therefore, the Fourier transformation is useful for calculating the edge magnetization.  
We write 
\begin{eqnarray}
d_{k_y} = \frac{1}{\sqrt{L_y}}\sum_{m=1}^{L_y} d_{(1,m)}e^{i k_y m}
\end{eqnarray}
and 
\begin{eqnarray}
c_{2\ell-1,k_y} = \frac{1}{\sqrt{L_y}}\sum_{m=1}^{L_y} c_{(2\ell-1,m)}e^{i k_y m}. \label{eq:c_fourier}
\end{eqnarray}
Then, the Zeeman term of Eq. (\ref{eq:ham_zeeman}) can be written as  
\begin{eqnarray}
H_{\mathrm{Z}}  = - i h \sum_{k_y} d_{k_y} c_{1, -k_y}, 
\label{eq:ham_zeeman_mom}
\end{eqnarray}
Notice that $\left(d_{k_y} \right)^\dagger = d_{-k_y}$ and  $\left(c_{2\ell-1, k_y} \right)^\dagger = c_{2\ell-1, -k_y}$.

However, it is still not so easy to diagonalize the Hamiltonian $H+H_{\mathrm{Z}}$ with a given finite magnetic field $h$
even if we impose the periodic boundary condition in the vertical direction instead of the open one. 
In the following, we will treat only the case with a sufficiently large system size and 
a sufficiently weak external magnetic field $h$. In this situation, it is enough to maximize 
the expectation value of the edge magnetization, or equivalently, 
to minimize the edge Zeeman energy in 
the sector of the quasi-degenerate ground states which is spanned by the states 
obtained by acting the edge mode operators,$ \gamma_{k_y}^{(2)}$ and $d_{k_y}$, on the fermion vacuum. 
To this end, we first rewrite $c_{1, k_y}$ in terms of the eigenmodes of $H$. 
As mentioned above, the Hamiltonian 
$H$ with the periodic boundary condition is exactly equal to 
the Hamiltonian (\ref{eq:hami}). 
Therefore, from (\ref{eq:def_alpha}), (\ref{eq:alpha_fourier}) and (\ref{eq:c_fourier}), one has 
\begin{eqnarray}
\alpha_{\ell, k_y} + \alpha^\dagger_{\ell, -k_y} = c_{2\ell-1, k_y},
\end{eqnarray}
which leads to
\begin{eqnarray}
\gamma^{(2)}_{k_y} = \sum_{\ell=1}^{L_x} u^{(2)}_{\ell,k_y} c_{2\ell-1, k_y},  \label{eq:anticommu}
\end{eqnarray}
from (\ref{eq:def_gamma}) and (\ref{eq:u2}).
Since $\{ c_{\ell,k_y}, c^\dagger_{\ell^\prime,k_y}  \} = 2 \delta_{\ell, \ell^\prime} $,
we have $\{c_{1,k_y}, \gamma^{(2) \dagger}_{k_y} \} = 2 u^{(2)}_{1,k_y}$ from Eq. (\ref{eq:anticommu}). 
Then, if we write $c_{1,k_y} $ as 
\begin{eqnarray}
c_{1,k_y} = W_{k_y} \gamma^{(2)}_{k_y} + \cdots, \label{eq:c_expansion} 
\end{eqnarray}
we obtain 
\begin{eqnarray}
W_{k_y} =\frac{1}{2} \{c_{1,k_y}, \gamma^{(2) \dagger}_{k_y} \} 
&=& \frac{1}{2} \{ c_{1,k_y}, \sum_{\ell=1}^{L_x} u^{(2) \ast}_{\ell,k_y} c^\dagger_{2\ell-1, k_y} \} \nonumber \\
&=&   u^{(2) \ast}_{1,k_y}. \label{eq:Wky}
\end{eqnarray}

Substituting Eqs. (\ref{eq:c_expansion}) and (\ref{eq:Wky}) into the Majorana representation of the magnetization per site in the $z$-direction, 
\begin{eqnarray}
\mu^z := \frac{i}{L_y} \sum_{k_y} d_{k_y} c_{1, -k_y}, 
\end{eqnarray}
we have 
\begin{eqnarray}
\mu^z &=& \frac{i}{L_y} \sum_{k_y} d_{k_y}\left[ u^{(2) \ast}_{1, -k_y} \gamma^{(2)}_{-k_y} + \cdots \right] \nonumber \\
 &=& \sum_{0 < k_y < \pi} \frac{i}{L_y N(k_y)} \left( 
 d_{k_y} \gamma^{(2) \dagger}_{k_y} +d^\dagger_{k_y} \gamma^{(2)}_{k_y}
 \right) + \cdots. \nonumber \\
\label{eq:ham_zeeman_mom2}
\end{eqnarray}
Here we have used
$u^{(2) \ast}_{\ell, -k_y} =u^{(2)}_{\ell, k_y}$, $u^{(2)}_{1,k_y} = \frac{1}{N(k_y)}$,
and $\gamma^{(2)}_{-k_y} = \gamma^{(2) \dagger}_{k_y}$,
which can be obtained from Eqs. (\ref{eq:def_gamma}) and (\ref{eq:u2}). 
The first term of Eq. (\ref{eq:ham_zeeman_mom2}) is a quadratic form of fermions
and can be easily diagonalized 
by introducing $p_{k_y} = \frac{1}{2} \left( d_{k_y} -i\gamma_{k_y}^{(2)} \right)$
and $q_{k_y} = \frac{1}{2} \left( d^{\dagger}_{k_y}  - i \gamma_{k_y}^{(2) \dagger} \right)$, as
\begin{eqnarray} 
&&  \sum_{0 < k_y < \pi} \frac{i} { L_y  N(k_y)} \left( 
 d_{k_y} \gamma^{(2) \dagger}_{k_y} +d_{-k_y} \gamma^{(2)}_{k_y}
 \right) \nonumber \\
&=& - 2 \sum_{0 < k_y < \pi}  \frac{1}{ L_y  N(k_y)} \left[ p^{\dagger}_{k_y} p_{k_y} + q^{\dagger}_{k_y} q_{k_y} -1 \right]  \nonumber \\
 &= & - \sum_{k_y}  \frac{1}{ L_y  N(k_y)} \left[ p^{\dagger}_{k_y} p_{k_y} + q^{\dagger}_{k_y} q_{k_y} -1 \right] , \nonumber \\ 
 \label{eq:zeeman_final}
\end{eqnarray}
whose expectation value is maximized by choosing $|\mathrm{GS}_{\mathrm{e}^\prime }  \rangle = | \tilde{0} , \Phi^{(0)}_{\mathrm{flux}} \rangle$,
where $\ket{\tilde{0}}$ is a vacuum for $p_{k_y}$ and $q_{k_y}$ that satisfies 
$p_{k_y} \ket{\tilde{0}} = q_{k_y}\ket{\tilde{0}} = 0$.
Here, we have used $p_{-k_y} = q_{k_y} $ and $q_{-k_y} = p_{k_y}$, in order to obtain the third line of Eq. (\ref{eq:zeeman_final}). 
The expectation value of the magnetization is given as 
\begin{eqnarray}
& & \langle \mathrm{GS}_{\mathrm{e}^\prime}|  \mu^z |\mathrm{GS}_{\mathrm{e}^\prime }   \rangle \nonumber \\
&\sim& - \sum_{k_y}  \frac{1}{ L_y  N(k_y)} \langle \mathrm{GS}_{\mathrm{e}^\prime } | \left[ p^{\dagger}_{k_y} p_{k_y} + q^{\dagger}_{k_y} q_{k_y} -1 \right] |\mathrm{GS}_{\mathrm{e}^\prime }\rangle \nonumber \\
&=&  \sum_{k_y} \frac{1}{L_y N(k_y)}    \nonumber \\
&=& U^{(2)}_{1}(1,1),
 \label{eq:mag2_final}
\end{eqnarray}
where, to obtain the final line of Eq. (\ref{eq:mag2_final}), we have used Eqs. (\ref{eq:u2}) and (\ref{eq:U}) .
In Eq. (\ref{eq:mag2_final}), we find that the edge magnetization is 
exactly equal to that in Eq. (\ref{eq:edgemag_final}).

\section{Majorana edge zero modes in bond-disordered systems \label{sec:appC}}
\begin{figure}[t]
\begin{center}
\includegraphics[width= 0.9\linewidth]{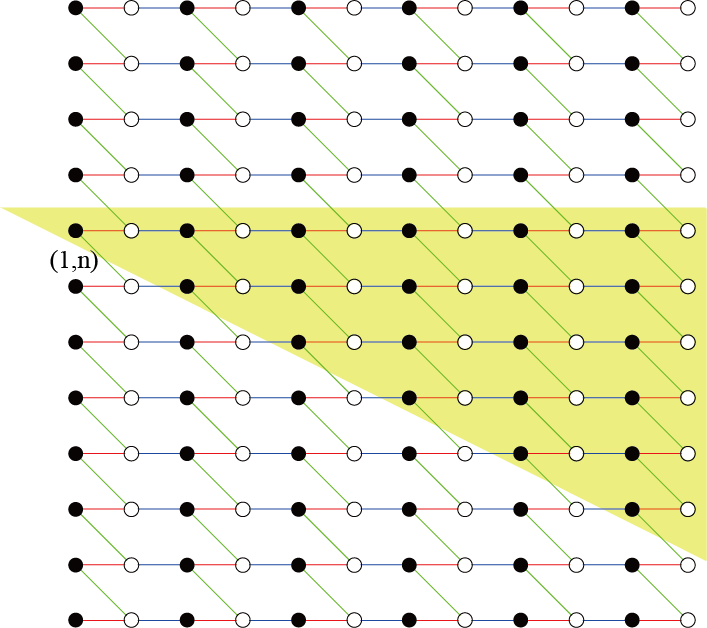}
\caption{
Schematic picture of the area where the zero mode $\gamma_n$ have a finite amplitude (a yellow shade). 
Note that the sites denoted by white circles do not have a finite amplitude. 
}
\label{Fig6}
\end{center}
\end{figure}
In this Appendix, we explain how to construct the edge zero modes of the disordered Hamiltonian of Eq. (\ref{eq:ham_disorder}). 

First, we rewrite the Hamiltonian in terms of the Majorana fermion, in exactly the same way as we have discussed in Sec. \ref{sec:model}.
Then, setting $d_i d_j = i$ for every $\langle i,j \rangle \in \mathcal{B}_z$, we obtain the Hamiltonian of the free Majorana fermion as 
 \begin{align}
 H  =&   i \sum_{\ell = 1}^{L_x} \sum_{m = 1}^{L_y} J_{x} [(2\ell-1,m),(2\ell,m)] c_{(2\ell -1,m)} c_{(2\ell,m)}  \notag \\
 +&   i\sum_{\ell = 1}^{L_x-1} \sum_{m = 1}^{L_y} J_{y} [(2\ell,m),(2\ell + 1,m)] c_{(2\ell ,m)} c_{(2\ell + 1,m)}   \notag \\
 +&  i\sum_{\ell = 1}^{L_x} \sum_{m = 1}^{L_y} J_{z} [(2\ell ,m),(2\ell -1,m + 1)]  c_{(2\ell,m)} c_{(2\ell-1,m+1)}. \label{eq:hami_disorder}
 \end{align}

For this Hamiltonian, we construct a set of the zero energy modes which are localized near the left edge. 
In the following, we assume that the length $L_x$ of the cylinder is large enough, so that the finite-size corrections 
about $L_x$ are exponentially small in the length $L_x$ and can be neglected.

Consider an operator,  
\begin{eqnarray}
\gamma_{n} = \sum_{\ell = 1}^{L_x} \sum_{m=1}^{L_y} 
U_{n} (\ell, m) c_{(2\ell-1,m)}, \label{eq:zeromode_disordered}
\end{eqnarray}
where $U_{n} (\ell, m)$ are real coefficients. This has the same form as in Eq. (\ref{eq:gamma_majorana}). 
We determine the coefficients $U_n(\ell,m)$ so that the operator $\gamma_n$ satisfies $[H,\gamma_n]=0$. 
Then, the mode $\gamma_n$ satisfies the zero energy condition. The normalization condition is given by 
\begin{equation}
1=\gamma_n^2=\sum_{\ell = 1}^{L_x} \sum_{m=1}^{L_y}|U_{n} (\ell, m)|^2. 
\end{equation}

By computing the commutation relation $[H,\gamma_n]=0$, one obtains
\begin{widetext}
\begin{eqnarray}
J_y [(2 \ell,m),(2\ell + 1,m)]U_{n} (\ell + 1,m) 
&=&{J_x[ (2\ell-1,m),( 2\ell,m) ]} U_{n} (\ell, m) \nonumber \\ 
&-&{J_z[ (2\ell,m),(2\ell-1,m+ 1) ]} U_{n} (\ell, m + 1). 
\end{eqnarray}
\end{widetext}
This implies that the coefficient $U_{n} (\ell + 1,m)$ is determined by the two coefficients, 
$U_{n} (\ell, m)$ and $U_{n} (\ell, m + 1)$, for a nonvanishing $J_y [(2 \ell,m),(2\ell + 1,m)]$. 
Therefore, when initial coefficients,  $U_n(1,m)$, for $m=1,2,\ldots,L_y$, are given,  
all the rest of the coefficients can be determined iteratively. In particular, when the exchange integrals 
satisfy 
\begin {widetext}
\begin{equation}
\left| J_x[ (2\ell-1,m),( 2\ell,m) ] \right|+ \left| J_z[ (2\ell,m),(2\ell-1,m + 1) \right| 
\le\kappa \left| J_y [(2 \ell,m),(2\ell + 1,m)] \right|
\end{equation}
\end{widetext}
with $\kappa\in(0,1)$ 
for all $\ell, m$, the coefficients, $U_{n}(\ell,m)$, decay exponentially in $\ell$. 

For the initial amplitudes, we choose 
\begin{eqnarray}
U_{n}(1, m )= \delta_{n,m}U_n(1,n)
 \label{eq:zeromode_condition1}
\end{eqnarray}
with a constant $U_n(1,n)\ne 0$, which is determined by the normalization condition. 
Then, clearly, one obtains the independent $L_y$ zero modes, $\gamma_{n}$, ($n=1, \cdots L_y$). 
We remark, however, that $\gamma_{n}$s are not orthogonal to each other in general,
i.e., $\left\{ \gamma_n, \gamma_{n^\prime} \right\} \neq 2 \delta_{n,n^\prime}$.
The orthogonal basis can be obtained by taking the linear combination of $\gamma_{n}$ properly. 
In Fig.~\ref{Fig6}, the yellow shaded region, which is triangular shaped, shows 
the area where the amplitude $U_{n} (\ell, m)$ may not be vanishing at each black circle site. 
All the amplitudes at the white circle sites are vanishing.

\end{document}